\documentclass[journal=accacs,manuscript=article,layout=twocolumn]{achemso}
\usepackage[version=3]{mhchem} 

\usepackage{multirow}
\usepackage{standalone}
\usepackage{mdframed}
\usepackage{rotating}
\usepackage{booktabs}
\usepackage{etoolbox}
\usepackage{enumitem}
\usepackage{subcaption}
\usepackage[nonumberlist]{glossaries}
\mciteErrorOnUnknownfalse
\usepackage{xcolor}
\usepackage{comment}
\usepackage{array}
\usepackage{hyperref}
\newcounter{magicrownumbers}
\preto\tabular{\setcounter{magicrownumbers}{0}}
\usepackage{todonotes}

\newcommand{\lz}[1]{\textcolor{blue}{LZ-comment: #1}}

\newcommand{\dataset}[1]{OC20}
\newcommand{\ocpurl}[0]{\url{http://opencatalystproject.org}}
\newcommand{\baselinesurl}[0]{\url{https://github.com/Open-Catalyst-Project/ocp}}
\newcommand{\dataseturl}[0]{\url{https://github.com/Open-Catalyst-Project/Open-Catalyst-Dataset}}

\makeglossaries
\newacronym{DFT}{DFT}{Density Functional Theory}
\newacronym{PES}{PES}{Potential Energy Surface}
\newacronym{QE}{QE}{Quantum Espresso}
\newacronym{VASP}{VASP}{Vienna Ab initio Simulation Package}
\newacronym{S2EF}{\textit{S2EF}}{Structure to Energy and Forces}
\newacronym{IS2RE}{\textit{IS2RE}}{Initial Structure to Relaxed Energy}
\newacronym{IS2RS}{\textit{IS2RS}}{Initial Structure to Relaxed Structure}
\newacronym{OC20}{OC20}{Open Catalyst 2020}
\newacronym{OOD}{OOD}{Out of Domain}
\newacronym{ML}{ML}{Machine Learning}
\newacronym{MD}{MD}{\textit{ab initio} Molecular Dynamics}
\glsaddall

\author{Lowik Chanussot}
\affiliation{Co-first Author}
\alsoaffiliation{Facebook AI Research}
\author{Abhishek Das}
\alsoaffiliation{Facebook AI Research}
\author{Siddharth Goyal}
\alsoaffiliation{Facebook AI Research}
\author{Thibaut Lavril}
\alsoaffiliation{Facebook AI Research}
\author{Muhammed Shuaibi}
\affiliation{Co-first Author}
\alsoaffiliation{Department of Chemical Engineering, Carnegie Mellon University}
\author{Morgane Riviere}
\affiliation{Facebook AI Research}
\author{Kevin Tran}
\author{Javier Heras-Domingo}
\affiliation{Department of Chemical Engineering, Carnegie Mellon University}
\author{Caleb Ho}
\affiliation{Facebook AI Research}
\author{Weihua Hu}
\affiliation{Computer Science Department, Stanford University}
\author{Aini Palizhati}
\affiliation{Department of Chemical Engineering, Carnegie Mellon University}
\author{Anuroop Sriram}
\affiliation{Facebook AI Research}
\author{Brandon Wood}
\affiliation{National Energy Research Scientific Computing Center (NERSC)}
\author{Junwoong Yoon}
\affiliation{Department of Chemical Engineering, Carnegie Mellon University}
\author{Devi Parikh}
\affiliation{Facebook AI Research}
\alsoaffiliation{School of Interactive Computing, Georgia Tech}
\author{C. Lawrence Zitnick}
\affiliation{Facebook AI Research}
\email{zitnick@fb.com}
\author{Zachary Ulissi}
\affiliation{Scott Institute for Energy Innovation, Carnegie Mellon University}
\alsoaffiliation{Department of Chemical Engineering, Carnegie Mellon University}
\email{zulissi@andrew.cmu.edu}

\title[]
  {The Open Catalyst 2020 (OC20) Dataset  and Community Challenges}

\abbreviations{IR,NMR,UV}

\let\oldmaketitle\maketitle
\let\maketitle\relax

\begin{document}



\twocolumn[
\begin{@twocolumnfalse}
\oldmaketitle

\begin{abstract}
\noindent Catalyst discovery and optimization is key to solving many societal and energy challenges including solar fuels synthesis, long-term energy storage, and renewable fertilizer production. Despite considerable effort by the catalysis community to apply machine learning models to the computational catalyst discovery process, it remains an open challenge to build models that can generalize across both elemental compositions of surfaces and adsorbate identity/configurations, perhaps because datasets have been smaller in catalysis than related fields. To address this we developed the OC20 dataset, consisting of 1,281,040 Density Functional Theory (DFT) relaxations ($\sim$264,890,000  single point evaluations) across a wide swath of materials, surfaces, and adsorbates (nitrogen, carbon, and oxygen chemistries). We supplemented this dataset with randomly perturbed structures, short timescale molecular dynamics, and electronic structure analyses. The dataset comprises three central tasks indicative of day-to-day catalyst modeling and comes with pre-defined train/validation/test splits to facilitate direct comparisons with future model development efforts. We applied three state-of-the-art graph neural network models (CGCNN, SchNet, DimeNet$++$) to each of these tasks as baseline demonstrations for the community to build on.  In almost every task, no upper limit on model size was identified, suggesting that even larger models are likely to improve on initial results. The dataset and baseline models are both provided as open resources, as well as a public leader board to encourage community contributions to solve these important tasks.
\end{abstract}
\end{@twocolumnfalse}
]

\clearpage

\section{Introduction}

Advancements to renewable energy processes are needed urgently to address climate change and energy scarcity around the world~\cite{Newell2020, domestic}. These include the generation of electricity through fuel cells, fuel generation from renewable resources, and the production of ammonia for fertilization. Catalysis plays a key role in each of these by enabling new reactions and improving process efficiencies.\cite{Norskov,She2017,Norskov2013}\ Unfortunately, discovering or optimizing catalysts remains a time-intensive process. The space of possible catalyst materials that can be synthesized or engineered is vast and modeling their full complexity under reaction conditions remains elusive. Simulation tools such as \gls{DFT}~\cite{Sholl2009}\ have greatly expanded our field's ability to develop reaction mechanisms for specific materials, rationalize experimental measurements, and suggest more active or selective structures for experimental testing. Despite steady growth in computing resources from Moore's law, the computational complexity of DFT remains a limiting factor in the large-scale exploration of new catalysts.\cite{doi:10.1021/acscatal.9b01234, doi:10.1021/acscatal.8b01708} Given its societal importance, finding computationally efficient methods for molecular simulations is of utmost necessity. One potentially promising approach is the use of efficient \gls{ML} models trained with data produced from computationally expensive models, such as DFT.

Indeed, the application of Artificial Intelligence and Machine Learning (AI/ML) to molecular simulations has increased in popularity recently, due to its ability to efficiently model complex functions in data-rich domains.  There have been a number of demonstrations from domain scientists for specific challenges such as reaction network elucidation\cite{ulissi2017address, li2019designing, gu2018thermochemistry}, thermochemistry prediction~\cite{Liu2019, Jirasek2020, kauwe2018machine, tran2020methods, Tran2018, gu2020practical, back2019convolutional, kim2019artificial,esterhuizen2020theory}, structure optimization~\cite{Sun2018, Li2019, Hayashi2020, schmidt2019recent, Aksoz2020}, accelerating individual calculations\cite{boes2017neural, khorshidi2016amp, peterson2016acceleration, sun2019toward}, and integration with characterization\cite{timoshenko2019inverting} (see recent reviews for a more thorough discussion \cite{kitchin2018machine, li2018toward,medford2018extracting,doi:10.1002/aic.16198,schlexer2019machine,li2017application,tabor2018accelerating,chen2020critical,artrith2019machine,gu2019machine,toyao2019machine,doi:10.1002/adma.201907865,SchlexerLamoureux2019,Takahashi2019}). Most of these tasks are variations on the same fundamental problem:  modeling heterogeneous catalysis. The dataset developed seeks to target a specific subclass of this problem, periodic slab models. Such modeling involves predicting the energy and forces of various configurations of adsorbate molecules at inorganic interfaces.

Unfortunately, modeling of heterogeneous catalysts entails all the known difficulties of modeling both organic and inorganic chemistry. In organic chemistry modeling involves an overwhelming space of molecules and reactions and many similar, low-energy conformers. Inorganic chemistry involves a  large diversity in elements, coordination environments, lattice structures, and long-range interactions. The result is a complex space of compositions and chemistries for which computationally efficient modeling methods are needed for thorough exploration.

A critical factor in building ML models is the data used for training. Despite the importance of heterogeneous catalysis, datasets for it remain smaller than those in other related fields\cite{Curtarolo2012, Kirklin2015} due to additional complexity and higher computational cost. Much of the progress in applying AI/ML in heterogeneous catalysis has been driven by increasingly large and diverse datasets of electronic structure calculations. In the past few years there has been a push towards larger datasets in catalysis, going from O(100)~\cite{calle2014fast, abild2007scaling, ma2017orbitalwise, noh2018active, andersen2019beyond} to O(1,000)~\cite{dickens2019electronic, li2017high, batchelor2019high} then O(100,000)~\cite{Mamun2019, tran2020methods, Winther2019} relaxations. Most focus on relaxed adsorption energies of simple adsorbates with smaller datasets of transition state calculations. State-of-the-art ML methods are still improving as data is added to these datasets, so there is no indication that we have saturated the performance of these models. Further, models trained on these datasets have shown limited ability to generalize, which suggests that the models are not yet learning fundamental physical representations. As has been shown in other ML tasks~\cite{deng2009imagenet,panayotov2015librispeech,antol2015vqa}, we expect that significantly larger datasets will lead to improved accuracy and better generalization.

\begin{figure}[t]
    \centering
    \includegraphics[width=0.5\textwidth]{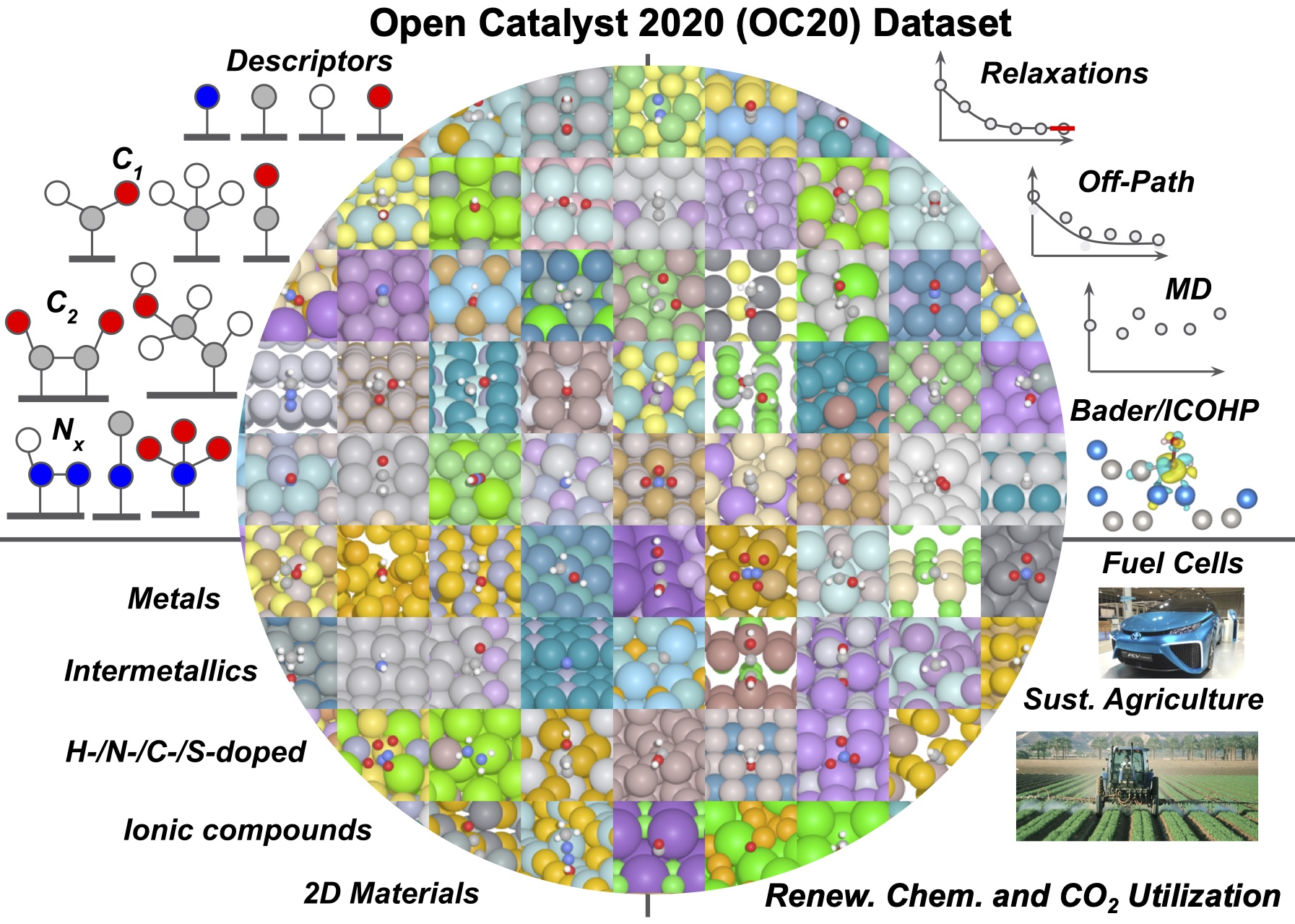}
    \caption{Adsorbates, materials, calculations,  and impact areas of the OC20 dataset. Images are a random sample of the dataset.}
    \label{fig:overview}
\end{figure}

In this paper, we present the \gls{OC20} dataset, (Figure~\ref{fig:overview})
which comprises over 1.2 million DFT relaxations of molecular adsorptions onto surfaces (\textit{ca.} 250 million single-point calculations) across a substantially larger structure and chemistry space than previously realized. We envision OC20 to serve as a crucial stepping stone in the development of ML models for practical catalysis applications. 

While a dataset of this magnitude will lead to significant improvements in ML models, this is still an extremely sparse sampling of all possibilities. We consider 82 different adsorbates (small adsorbates, C\textsubscript{1}/C\textsubscript{2} compounds, and N/O-containing intermediates) that are relevant for renewable energy and environmental applications. Relaxations are performed on randomly sampled low-Miller-index facets of stable materials from the Materials Project \cite{Jain2013}, resulting in surfaces from 55 different elements and mixtures thereof. For each of the calculations, we include relaxation trajectories, Bader charges, and LOBSTER~\cite{henkelman2006fast, bader1990atoms}-calculated orbital information. To aid in training more robust models, we additionally compute short, high-temperature \gls{MD} trajectories on a randomly sampled subset of the relaxed states. We also randomly perturb the atomic positions in a subset of the structures along the relaxation pathways and perform single point DFT calculations for these perturbed/rattled structures. We recognize that OC20 addresses a simplified version of heterogeneous catalysis - single adsorbates on idealized structures. Although useful as a first step to informing reaction pathways, the reality involves a number of additional complexities that impact catalyst performance, including reaction conditions, solvation effects, kinetics, etc. While we believe OC20's approximations to be a reliable step forward, it is important to understand the limits of models developed from this dataset. Future work that incorporates more of the complexities mentioned will undoubtedly benefit from the developments related to OC20. The dataset is publicly available at \ocpurl{}. We also plan to upload the dataset to other open systems (e.g. NOMAD or Zenodo) for long-term availability.

In addition to generating and sharing the dataset, we propose three related domain challenges as an open competition:  (1) predict the energy and force for a given state, (2) predict a nearby relaxed state given an initial starting state, and (3) predict the relaxed adsorption energy given an initial state. The dataset is split into train/validation/test splits indicative of common situations in catalysis:  predicting these properties for a previously unseen adsorbate, for a previously unseen crystal structure or composition, or both. To boot-strap research and the competition, we also provide an open software repository (\baselinesurl{}) containing a set of baseline models, data loaders, and training scripts for each of these tasks. While we focus on a subset of tasks, we believe that models capable of solving these tasks on the OC20 dataset will also be able to address a large number of related catalysis problems.
\begin{figure*}[ht]
    \centering
    \includegraphics[width=\textwidth]{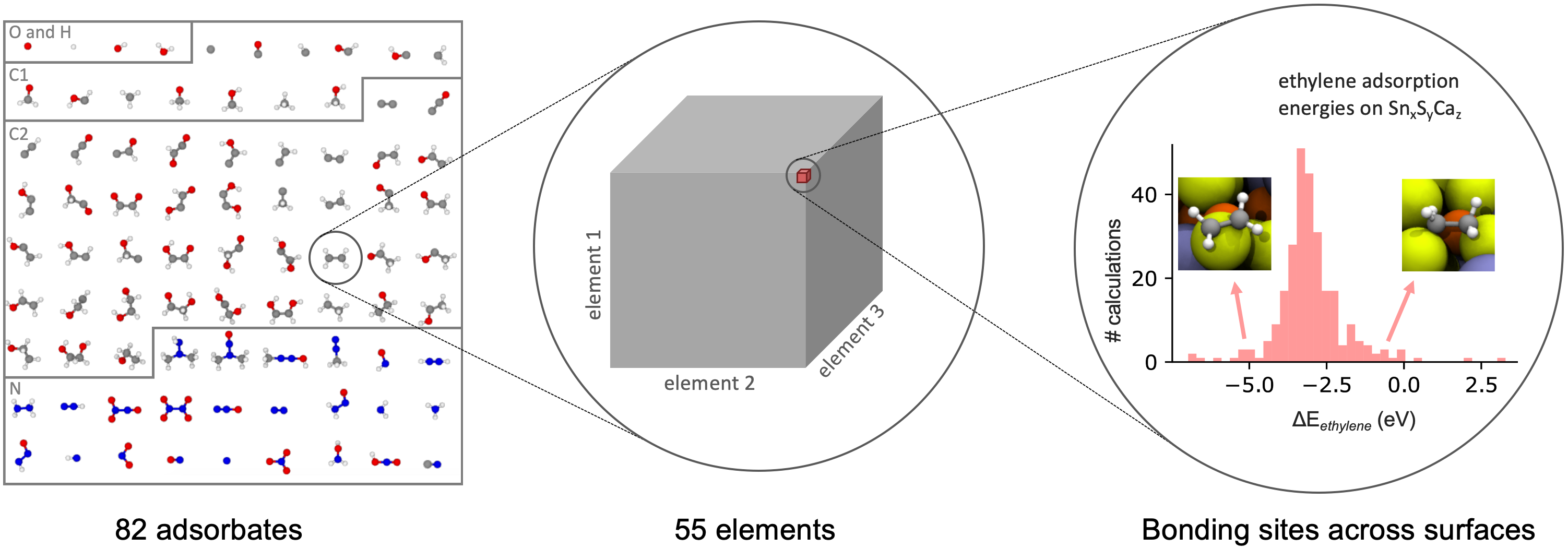}
    \caption{The adsorbates used to generate the Open Catalyst Dataset contain oxygen, hydrogen, C\textsubscript{1}, C\textsubscript{2}, and nitrogen molecules useful for renewable energy applications. Adsorbates that contain both carbon and nitrogen were counted both as C\textsubscript{X} adsorbates and as nitrogen-containing adsorbates. For each adsorbate, up to 55$^3$ different catalyst compositions were considered, with up to dozens of adsorption energy calculations per adsorbate-composition pairing.}
    \label{fig:adsorbates}
\end{figure*}

\section{Tasks}

Our goal is to improve the efficiency with which inorganic and organic interfaces can be simulated for use in catalysis. Since the primary computational bottlenecks are the DFT calculations used to compute a structure's forces and energy, we focus on the general challenge of efficient DFT approximation. We focus on structure relaxation -- a fundamental calculation in catalysis used in determining a structure's activity and selectivity. We define three related tasks, in that success in one task may aid other tasks.  These are not the only possibilities for this dataset, and future tasks may be added with additional data generation and input from the community.

In all our tasks, the structure contains a surface and adsorbate. The surface is defined by a unit cell that is periodic in all directions with a vacuum layer of at least 20\AA{} applied in the \textbf{z} direction. Initial structures are heuristically determined. Ground truth data is computed for all tasks using DFT. Dataset details and evaluation metrics are provided in following sections.

\textbf{\gls{S2EF}} is to take the positions of the atoms as input and predict the energy and per-atom forces as calculated by \gls{DFT}. For the purposes of this manuscript, energy refers to adsorption energy unless otherwise noted. The adsorption energy is defined as the energy of the combined surface and adsorbate system (relaxed or not) minus the energy of the relaxed slab and the relaxed gas phase adsorbate molecule. The force is defined as the negative gradient of the energy with respect to the atomic positions.

This is our most general task and has the broadest applicability across catalysis and related fields.
It is essentially identical to existing challenges in developing machine learning potentials~\cite{von2020exploring}. However, the inclusion of both inorganic and organic materials and the dataset size make this challenge unique.

\textbf{\gls{IS2RS}} takes as input an initial structure and predicts the atomic positions in their final, relaxed state. Traditional relaxations are performed through an iterative process that estimates the atomic forces using DFT, which are in turn used to update atom positions until convergence. This very computationally expensive process typically requires hundreds of DFT calculations to converge.

If the \gls{IS2RS} task is approached using ML approximations of DFT to estimate atomic forces (\gls{S2EF} task), evaluation on the \gls{IS2RS} task may help determine whether models built for \gls{S2EF} are sufficiently accurate for practical applications. Alternatively, it may be possible to predict the relaxed structure directly, without estimating a structure's energy or forces (Figure \ref{fig:relaxations}(B)), as many of the changes during relaxation (say due to particular initial guess strategies) are systematic. These direct IS2RS approaches may lead to even further improvements in computational efficiency.

\textbf{\gls{IS2RE}} task is to take the initial structure as input and predict the structure's energy in the relaxed state. This is the most common task in catalysis, as the relaxed energies are often correlated with catalyst activity and selectivity, and the energies are important parameters for detailed microkinetic models. Similar to \gls{IS2RS}, this task may be approached by estimating the relaxed structure and energy by iteratively applying \gls{S2EF}, or by directly regressing the energy from the initial structure without estimating the intermediate or relaxed structures.

\section{The OC20 Dataset}

The \gls{OC20} dataset is constructed to provide both training and evaluation data for our three previously defined tasks involving DFT approximation and structure relaxation. Modern machine learning models, especially those employing deep learning, require sufficiently large datasets to learn accurate models. For training, we provide 640,081 relaxations across a wide variety of surfaces and adsorbates. The intermediate structures and their corresponding energy and forces are provided for each relaxation resulting in over 133 million training structures. To potentially aid in training and to provide additional information for the catalysis community, we performed DFT calculations on rattled and \textit{ab initio} Molecular Dynamics (MD) data. We also computed Bader charges and LOBSTER analyses (over 1.8 million examples each) as these computed properties may be useful for models by explaining why the energies are what they are.

\subsection{Dataset Generation}

The dataset is constructed in four stages: 1) adsorbate selection, 2) surface selection, 3) initial structure generation, and 4) structure relaxation. We describe each of these four stages in turn, followed by a description of the additional data provided with the main dataset. All source code to generate the configurations are provided in the Open Catalyst Dataset repository (\dataseturl{}).

\subsubsection{Adsorbate Selection}

Adsorbates are sampled randomly from a set of 82 molecules that are chosen for their utility to renewable energy applications.  As shown in Figure~\ref{fig:adsorbates}, this includes adsorbates that contain only oxygen or hydrogen, C\textsubscript{1} molecules, C\textsubscript{2} molecules, and nitrogen-containing molecules. We enumerated the oxygen and hydrogen molecules for their ubiquitous presence in water-solvated electrochemical reactions. C\textsubscript{1} and C\textsubscript{2} molecules are important for solar fuel synthesis, while nitrogen-containing molecules have applicability in solar fuel and solar chemical synthesis. Note that some of the C\textsubscript{2} molecules have two binding sites; we refer to these as bidentate adsorbates. The list of all 82 adsorbates is provided in the Supplementary Information.

\subsubsection{Surface Selection}

Surfaces are sampled in three stages. First, the number of elements is selected with a 5\% chance of choosing a unary material, 65\% chance for a binary material, and a 30\% chance for a ternary material. Greater emphasis is given to binary and ternary materials because these sets contain a wider variety of understudied materials. Next, a stable bulk material is randomly selected from the 11,451 materials in the Materials Project\cite{Jain2013} with the number of elements chosen in the first step. Finally, all symmetrically distinct surfaces from the material with Miller indices less than or equal to 2 are enumerated, including possibilities for different absolute positions of surface plane. From this list of surfaces one is randomly selected. The surface atoms were replicated to a depth of at least 7 \AA{} and a width of at least 8 \AA{}.

Pymatgen\cite{Ong2013} was used to search over all bulk materials in the Materials Project with non-positive formation energies and energies-above-lower-hulls of at most 0.1 eV/atom. The enumeration of symmetrically distinct surfaces was also performed using pymatgen\cite{Ong2013}. Elements for the bulk materials were chosen from a set of 55 elements comprising reactive nonmetals, alkali metals, alkaline earth metals, metalloids, transition metals, and post-transition metals.

Note that DFT was used to re-optimize the bulk structures prior to surface enumeration to ensure differences between the DFT settings used in the Materials Project and OC20 did not induce unintended stress or strain effects. Any bulks that we could not successfully relax were omitted from this dataset.

\begin{table*}[ht]
    \begin{tabular}{l|r|r|r|r|r}
        \toprule
        Task & \multicolumn{1}{|c|}{Train} & In Domain & OOD Adsorbate & OOD Catalyst   & OOD Both \\
        \midrule 
        S2EF & 133,934,018 & 987,036 & 999,838 & 987,343 &  997,922 \\
        IS2RS & 460,328 & 24,733 & 24,961 & 24,738 & 24,971 \\
        IS2RE & 460,328 & 24,733 & 24,961 & 24,738 & 24,971 \\
        \bottomrule
    \end{tabular}
    \caption{Size of train/validation splits (number of structures for \gls{S2EF} and initial structures for \gls{IS2RS} and \gls{IS2RE}). The structures for \gls{S2EF} are sampled from 640,081 relaxations for train, and from 30k-70k relaxations for each validation and test split. Subsplits of validation and test are roughly the same size, but are exclusive of each other. Subsplits include sampling from the same distribution as training (In Domain), unseen adsorbates (\gls{OOD} Adsorbate), unseen element compositions for catalysts (\gls{OOD} Catalyst), and unseen adsorbates and catalysts (\gls{OOD} Both). Test sizes are similar. \label{tab:splits}}
\end{table*}

\subsubsection{Initial Structure Generation}
The initial structures are generated by placing the selected adsorbates on the selected surfaces using CatKit~\cite{boes2019graph} and the atomic simulation environment (ASE)~\cite{ase_Hjorth_Larsen_2017}. Surface atoms are identified by their positions above the center-of-mass, their z-distance within 2 \AA{} of the upper-most atom, and by their under-coordination relative to the bulk atoms. Atomic coordination environments were calculated using pymatgen's Voronoi tesselation algorithm~\cite{Ong2013}. Next, we manually tagged the adsorbates' binding atoms for both mono- and bi-dentate adsorbates. Finally, we gave the surface structure, adsorbate, the identified surface atoms, and identified adsorbate binding sites to CatKit.\cite{boes2019graph} CatKit used this information to enumerate a list of symmetrically distinct adsorption sites along with suggested per-site orientations for the adsorbates. From this list, an adsorption configuration is randomly selected. The sites selected are not necessarily the most stable adsorption site on each surface. Since one of our goals is to calculate adsorption energies, we generate two sets of inputs for each system:  (1) the adsorbate placed over the catalyst atoms, and (2) just the catalyst atoms without the adsorbate. This resulted in a total of 1,919,165 and 616,124 unique inputs for (1) and (2), respectively, which were later filtered and segregated into suitable train, validation, and test validation splits as described later in this section.

\subsubsection{Structure Relaxation}
All structure relaxations were performed using the Vienna Ab Initio simulation Package (VASP)~\cite{Kresse1994, Kresse1996, Kresse1996a, vasp-license, kresse1999ultrasoft} until all per-atom forces are less than 0.03 eV/\AA.  Calculations were allowed up to 144 hours (12 cores) for the relaxation. Systems that timed out before reaching the specified force threshold were set aside for the S2EF task. All intermediate structures, energies, and forces are stored for future training and evaluation.  During the relaxations only adsorbate and surface atoms (as defined during the generation above) were allowed to move; subsurface atoms were maintained at fixed positions. This was done to avoid unrealistic structure deformations and to simulate a semi-infinite condition with bulk material far below the catalyst surface. Given the intended scale of OC20, the careful consideration of DFT settings was a non-trivial challenge. Relaxations generally followed previous high-throughput catalysis efforts with reasonable trade-offs between accuracy for surface chemistry and computational cost\cite{Tran2018} (VASP~\cite{Kresse1994, Kresse1996, Kresse1996a, vasp-license, kresse1999ultrasoft}, RPBE\cite{perdew1996generalized}, no spin polarization, etc). The choices made for DFT were a result of several important considerations: ensuring calculations were representative, concerns associated with inconsistent cutoffs/settings, and representative of typical numerical/convergence issues the computational chemistry field faces. The assumptions made were necessary to achieve the dataset's scale. Detecting small numerical or convergence errors is a non-trivial problem that could be improved with this dataset. Most importantly, we anticipate models and methods that solve the S2EF, IS2RE, or IS2RS tasks for this dataset are very likely to solve future challenges for future surface science datasets with different DFT modeling choices.

System DFT energies were referenced to represent adsorption energies. Adsorption energies were calculated according to the Equation below, where $E_{sys}$ is the DFT energy of the combined surface (i.e. slab) and adsorbate — this energy can be from both relaxed and intermediate structures. The reference energies for each system, $E_{slab}$ and $E_{gas}$ are the DFT energy of the relaxed surface and adsorbate molecule respectively. The value of $E_{gas}$ for each adsorbate was computed as a linear combination of N$_2$, H$_2$O, CO, and H$_2$ resulting in the atomic energies found in the supplementary.
\begin{align*}
    E_{ad} = E_{sys} – E_{slab} – E_{gas}
\end{align*}
Resulting trajectories were further analyzed for per-atom force criterion, numerical issues, or catastrophic reconstructions as described below in the Train, Validation, and Test Splits section.

\subsubsection{MD and Rattled Calculations}
The intermediate structures from the relaxations may result in a dataset biased towards structures with lower energies. To learn robust models, training samples with higher forces and greater configurational diversity may be needed. We adopted two strategies for generating additional training data:  (1) partial MD in VASP~\cite{Kresse1994, Kresse1996, Kresse1996a, vasp-license, kresse1999ultrasoft} and (2) normally-distributed random position perturbation methods colloquially known in molecular simulations as ``rattling.''

MD calculations simulate the atomic interactions when heat is added to the system. Partial MD calculations were carried out on previously relaxed structures with random initial velocities generated from a Maxwell-Boltzmann distribution at a temperature of 900 K. We integrated the MD trajectories over 80 fs or 320 fs with integration steps of 2 fs in the NVE ensemble. Time-scales were selected to allow systems to explore local configurations while minding computational costs. 

To diversify the distribution of single-point structures in the dataset, we ``rattled'' some of the structures by adding random displacements to the atomic positions with ASE~\cite{ase_Hjorth_Larsen_2017}. For each relaxation, 20\% of the images in the trajectories were selected for rattling. The atomic displacements were sampled from a heuristically-generated normal distribution with a $\mu=0$ and $\sigma=0.05$. Single point DFT calculations were then performed on the rattled structures.

Similar to the relaxations, only the top surface atom layers were allowed to move in both the MD and rattled calculations with the rest of the atom positions held fixed. All calculations were performed at the same theoretical level and energy/forces convergence criteria as in the relaxation calculations. Approximately 950 thousand MD (\textit{ca.} 64 million single-point energies/forces) and 30 million rattled calculations were carried out.

\subsubsection{Bader Charges and LOBSTER Analyses}
We performed electronic structure calculations for general use by the catalysis research field. These calculations (i.e., Bader charges~\cite{tang2009grid,sanville2007improved, henkelman2006fast} and LOBSTER~\cite{nelson2020lobster, deringer2011crystal} analyses) were carried out on relaxed structures and also on randomly selected snapshots from both MD and rattled trajectories. Bader charge analyses provides charge density maxima at each atomic center and the Bader volume for each atom through the zero-flux partitioning method~\cite{bader1990atoms}. LOBSTER enables chemical-bonding analysis based on periodic DFT outputs~\cite{nelson2020lobster}. LOBSTER calculates atom-projected densities of states (pDOS) or projected crystal orbital Hamilton population (pCOHP) curves, among others. Literature has demonstrated that such electronic structure information can provide valuable insights to the theoretical and the ML communities~\cite{gong2019predicting,nagai2020completing,chandrasekaran2019solving}.

\subsubsection{Dataset profile}
Approximately 872,000 adsorption energies were calculated successfully. Of these, 3.7\% were calculations on unary catalysts; 61.4\% were on binaries; and 34.9\% were on ternaries. Among these calculations, 28.9\% of them had reactive nonmetal elements in the catalyst; 8.1\% of them had alkali metals; 10.2\% had alkaline earth metals; 26.4\% had metalloids; 81.3\% had transition metals; and 37.2\% had post-transition metals. Considering adsorbates:  6.6\% of the calculations had adsorbates containing only oxygen or hydrogen; 25.2\% of the calculations had C\textsubscript{1} adsorbates; 44.4\% had C\textsubscript{2} adsorbates; and 29.0\% had nitrogen-containing adsorbates.

Despite this dataset's large size compared to previous catalytic datasets, it still very small in comparison to the number of potential calculations. Of the ${55 \choose 3}+{55 \choose 2}+{55 \choose 1}=27,775$ possible compositions, only 5,243 (18.9\%) of them were successfully sampled here. Of the compositions sampled, there were an average of 249 successful adsorption calculations for each. Additionally:  if we compare the number of sites we sampled here to rough estimates of the number of sites we could have sampled given our constraints on adsorbates, surfaces, and bulks, then we find that we performed \textit{ca.} 0.07\% of the possible calculations. This severe sparsity in the data compared to its large scale emphasizes the need for surrogate models.

\subsection{Train, Validation and Test Splits}

We split our dataset into training, validation, and testing sets. The training set is used to learn model parameters; the validation set is used to tune model hyperparameters and to perform ablation studies; and the test set is used to report model performance.

A careful choice of validation and test splits can help evaluate a model's performance on both interpolative and extrapolative tasks. Interpolative evaluation tests the ability to model variations of the training data, and is performed by sampling examples from the same distribution as the training dataset. Extrapolative evaluation tests a model's performance on unseen tasks, e.g., new materials or adsorbates. In the context of catalytic development, we strive to extrapolate beyond data we have already seen so that we can discover new materials and search spaces~\cite{Kim2019, Meredig2018}.

We explore extrapolation along two dimensions; new adsorbates and new catalyst compositions. Adsorbate extrapolation is performed by holding out 14 adsorbates from the training dataset sampled from all types (O, H, C1, C2, and N) of adsorbates. Similarly for catalyst compositions, a subset of element combinations for catalysts is held out from the training dataset. These were sampled from the 1,485 binary and 26,235 ternary material combinations of the 55 elements used in the dataset. No surfaces with unary materials are in the extrapolative subsplits for validation and testing. A full list of the adsorbates materials in train and validation splits are in the SI. 


We used four subsplits for each of the validation and test sets by considering all combinations of potential extrapolations (Table~\ref{tab:splits}). These include In-Domain (sampled from the training distribution), Out-of-Domain Adsorbate (OOD Adsorbate), OOD Catalyst, and OOD Both (both unseen adsorbate and unseen catalyst compositions). As shown in Table~\ref{tab:splits}, each subsplits in validation and testing contains \textit{ca.} 25,000 relaxations. For the \gls{S2EF} task we randomly select a one million structure subset from the relaxations in each subsplit. Note that the extrapolative subsplits of our validation set are completely exclusive to the extrapolative subsplits in the test set, e.g., the adsorbates in the validation adsorbate subsplit are unique from the adsorbates in the test adsorbate subsplit. This helps ensure overfitting to the test set does not occur during hyperparameter tuning on the validation set.

\begin{figure*}[ht]
    \centering
    \includegraphics[width=\textwidth]{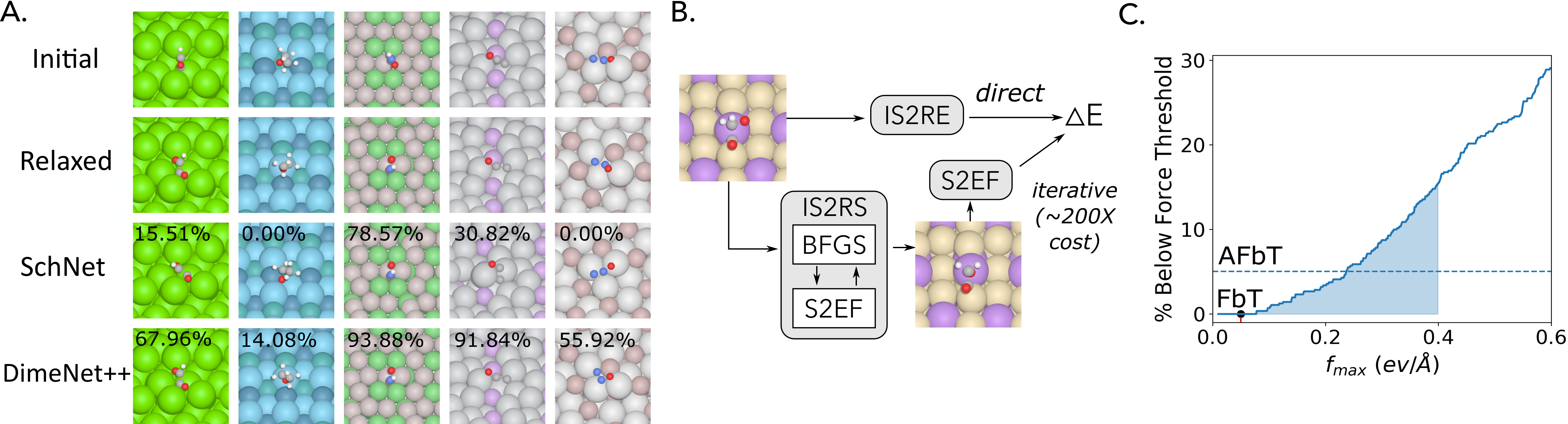}
    \caption{Demonstration of baselines SchNet and DimeNet$++$ models for solving the \gls{IS2RE}, \gls{S2EF}, and \gls{IS2RS} tasks and the inter-relationships. (A) Snapshots of five representative initial adsorbate configurations before DFT relaxations, the same adsorbates after DFT relaxation, and the relaxed structures as relaxed by SchNet and DimeNet$++$ after fitting the \gls{S2EF} task. ADwT metrics are overlaid on the model snapshots. (B) Three ways to predict the relaxed energy: directly through \gls{IS2RE}, indirectly through \gls{IS2RS}, and confirmation of the relaxed structure with a single DFT single-point. (C) SchNet force-only performance as characterized by the percentage of structures within the desired max force threshold of 0.05 eV/\AA (FbT) and average percentage of force below threshold (AFbT) of 0.4 eV/\AA (shaded area). \label{fig:relaxations}}
\end{figure*}

\section{Baseline GNN Models}
\label{sec:baselines}

We evaluate our tasks using a set of baseline models that are representative of the current
state-of-the-art. The set of models we evaluate is by no means comprehensive, but they demonstrate what is feasible with current models. Code and pretrained models for our baseline ML approaches implemented in PyTorch Geometric~\cite{fey2019fast,paszke2019pytorch} are publicly available at the Open Catalyst Project (\ocpurl{}).

Our baseline ML approaches are all based on Graph Neural Networks (GNNs)~\cite{hamilton2017representation} that operate over a graph structure containing nodes and edges. In our domain, the nodes represent atoms and edges represent the relationship between neighboring atoms.  At each node, an atom embedding is iteratively updated based on messages passed along the edges. During this message-passing phase, GNNs employ neural networks to learn the atomic representations~\cite{Behler2016, Bartok2010}, and unlike traditional descriptor-based models do not require hand-crafting. Node embeddings are initialized based on the atom's properties, such as their atomic number, group number, electronegativity, atomic volume, etc.~\cite{xie2018crystal} Outputs for the GNN may be computed from individual node (atom) embeddings for node-specific information (per-atom forces), or over the pooled node embeddings for system outputs (structure energy).

We benchmark three recent GNN methods: Crystal Graph Convolutional Neural
Network (CGCNN)~\cite{xie2018crystal}, SchNet~\cite{schutt2017schnet} and DimeNet$++$~\cite{klicpera2020fast, klicpera2020directional}. CGCNN is one of the first approaches to use GNNs on periodic crystal systems and uses a diverse set of features as input to the node embeddings. The original model encoded edge information using the discretized distances between atoms. SchNet proposed using continuous edge filters, which allows for the computation of per-atom forces through partial derivatives of the structure's energy with respect to the atom positions. To allow CGCNN to compute per-atom forces in the same manner, we updated the distance encoding to use gaussian basis functions but without the envelope distance function used in SchNet in our experiments. Finally, to not only encode distance information but also angular information between triplets of atoms, DimeNet introduced the use of directional message passing. DimeNet$++$, an extension to DimeNet, replaces the Bilinear layer with a Hadamard product and additional multilayer perceptrons; providing reported speed improvements of 8x and a 10\% accuracy boost on QM9~\cite{ramakrishnan2014quantum}.

For all approaches, graph edges were determined by a nearest neighbor search limited by a cutoff radius of 6\AA, retaining up to the 50 nearest neighbors. When computing distances, periodic boundary conditions were taken into consideration. Atoms were tagged as three types, slab (fixed), surface (free), and adsorbate (free), to allow loss functions to emphasize free atoms over fixed atoms. The number of hidden channels is 128, 1024, 192 for CGCNN, SchNet and DimeNet$++$ respectively unless stated otherwise; resulting in 3.6 million (CGCNN), 7.4 million (SchNet) and 1.8 million (DimeNet++) parameters. Model sizes were chosen so that runtimes were roughly equivalent. Note the size of the models was increased from their original implementations to account for \gls{OC20}'s larger size. Model hyperparameters and additional modifications can be found in the supplementary.

Since both the computed energies and forces are evaluated, the baseline loss function~\cite{klicpera2020directional, khorshidi2016amp} uses the following form:
\begin{align*}
    \mathcal{L}=\lambda_E\sum_i |E_i-E_i^{DFT}|\\ +\lambda_F\sum_{i,j}\frac{1}{N_i} |F_{i,j}-F_{i,j}^{DFT}|,
\end{align*} where $\lambda_E$ and $\lambda_F$ are empirical parameters, $E_i$ is the energy of image $i$, and $F_{i,j}$ is the force of the $j$th free atom in image $i$, and $N_i$ is the number of free atoms in image $i$. For the \gls{IS2RE} task, in which only the energy is evaluated, only the first term of the loss function is used ($\lambda_F = 0$).

All of the models are ML-based as there are currently no physical models that operate over such a large composition space with reasonable accuracy and elemental parameterizations. In particular, the recently developed GFN0-xTB method \cite{pracht2019robust} is parameterized for all of the elements in this dataset and is fast enough (approx 1,000X faster than DFT) to compete on these benchmarks and preliminary results are reported in the SI. However, since the method was not fit for inorganic surfaces and the xTB code \cite{bannwarth2020extended} is still under active development for periodic boundary conditions, the results were excluded from the summaries here. We hope that the release of our dataset will inspire future efforts on parameterizating tight-binding DFT codes or reactive force field methods for these materials.

\section{Experiments}
We begin by describing the metrics used to evaluate our three tasks, followed by the results of our baseline models.

\subsection{Evaluation Metrics}

For each task, we define evaluation metrics to track the progress in the field, as well as to measure the practical utility of the approaches. All ground truth values are computed using DFT. Our evaluation metrics are as follows:

{\bf \gls{S2EF}:} The \gls{S2EF} task has three metrics: the Mean Absolute Error (MAE) for energy, MAE for forces on free atoms and a combined metric. Our combined metric, Energy and Forces within Threshold (EFwT), is designed to measure the practical usefulness of a model for replacing DFT by evaluating whether both the computed energy and forces are close to the ground truth.
\begin{itemize}[leftmargin=*]
    \item[] {\bf Energy MAE:} Mean Absolute Error between the computed energy and the ground truth energy.
    \item[] {\bf Force MAE:} Mean Absolute Error between the computed per-atom forces and the ground truth forces. Errors are only computed for free catalyst and adsorbate atoms.
    \item[] {\bf Force cosine:} Mean cosine of the angle between the computed per-atom forces and the ground-truth forces. Similar to MAE, these are only computed for free atoms.
    \item[] {\bf EFwT:} The percentage of structures in which the computed energy is within $\epsilon = 0.02$ eV of the ground truth energy, and the maximum error in per-atom forces is below $\alpha = 0.03$ eV/{\AA}. Both these criteria must be met for the structure to be labeled as ``correct''.
\end{itemize}

{\bf \gls{IS2RS}:} Several methods exist for determining the accuracy of relaxed structures predicted by ML models. The simplest is to measure the distance between the predicted 3D positions of the atoms and those of the ground truth. However, small changes in position can lead to significant changes in the per-atom forces and a structure's energy. For this reason, a better measure of a proposed relaxed structure is the magnitude of its per-atom forces as measured by a single point DFT calculation. If the proposed relaxed structure represents a true local energy minimum, the forces should be close to zero.

\begin{itemize}[leftmargin=*]
    \item[] {\bf ADwT:} The Average DwT (Distance within Threshold) across thresholds ranging from $\beta =0.01${\AA} to $\beta=0.5${\AA} in increments of 0.001{\AA}. DwT is computed as the percentage of structures with an atom position MAE below the threshold. MAE is only computed for free catalyst and adsorbate atom positions while taking into account periodic boundary conditions. We use ADwT as opposed to the MAE on 3D atom positions, since ADwT is robust to outliers and better indicates the percentage of relaxations that are likely to be successful.
    \item[] {\bf FbT:} The percentage of relaxed structures with maximum DFT calculated per-atom force magnitudes below a threshold of $\alpha=0.05$ eV/{\AA}. Force magnitudes of only free catalyst and adsorbate atoms are used. A value of $\alpha=0.05$ eV/{\AA} represents a practical threshold by which DFT relaxations are commonly assumed to have converged. To ensure that the ML relaxations find a relaxed structure that isn't significantly different from the ground truth relaxed structures, e.g., the adsorbate moves to a different binding site, an additional filtering step is applied. We filter on the atom position MAE (free catalyst and adsorbate atoms) with a threshold of $\beta=0.5${\AA}. Thus, to be considered correct, a relaxed structure must meet both the FbT and the DwT criterion.
    \item[] {\bf AFbT:} The Average FbT (Forces below Threshold) over a range of thresholds ranging from $\alpha = 0.01$ eV/{\AA} to $\alpha = 0.4$ eV/{\AA} in increments of 0.001 eV/{\AA}, Figure \ref{fig:relaxations}(C). This metric measures progress over a wider range of thresholds, which may be important for early algorithm development that may need thresholds more lenient than $\alpha = 0.05$ eV/{\AA} to see improvement. Similar to FbT, the relaxed structures must also meet the same DwT criterion with $\beta=0.5${\AA}.
\end{itemize}

Note that FbT and AFbT require the computation of single point DFT calculations, which are computationally expensive. For this reason, a random subset of 500 relaxed structures are chosen from the validation and test set splits (2000 total for each) for evaluating these metrics. If a DFT calculation fails to converge within 60 electronic steps or a wall time of 2 hrs, the system is assumed to be incorrect with forces beyond the thresholds for both FbT and AFbT.

\setlength{\tabcolsep}{3pt}
\begin{table}[h!]
    \begin{center}
        \resizebox{\columnwidth}{!}{
            \begin{tabular}{l cccc }
            & \multicolumn{4}{c}{\gls{S2EF} Test}  \\
            \midrule
            Model  & ID &  OOD Ads & OOD Cat & OOD Both \\
            \midrule
            & \multicolumn{4}{c}{Energy MAE [eV] $\downarrow$} \\
            Median baseline  & $2.0596$ & $2.4188$ & $2.0110$ & $2.5460$ \\
            CGCNN~\cite{xie2018crystal} & $0.5105$ & $0.6321$ & $0.5202$ & $0.7681$ \\
            SchNet~\cite{schutt2017schnet} & $0.4421$ & $0.4858$ & $0.5279$ & $0.7057$ \\
            SchNet~\cite{schutt2017schnet} -- force-only & $34.0689$ & $33.7670$ & $35.2701$ & $38.4607$ \\
            SchNet~\cite{schutt2017schnet} -- energy-only & $0.3975$ & $0.4533$ & $0.5626$ & $0.7241$ \\
            DimeNet$++$~\cite{klicpera2020directional,klicpera2020fast} & $0.4579$ & $0.4701$ & $0.5056$ & $0.6489$ \\
            DimeNet$++$~\cite{klicpera2020directional,klicpera2020fast} -- force-only & $28.2214$ & $28.9404$ & $28.8636$ & $34.9118$ \\
            DimeNet$++$~\cite{klicpera2020directional,klicpera2020fast} -- energy-only & $0.3585$ & $0.4022$ & $0.5041$ & $0.6549$ \\
            DimeNet$++$~\cite{klicpera2020directional,klicpera2020fast}-Large -- force-only & $29.3504$ & $30.0338$ & $30.0074$ & $36.7665$ \\
            \midrule
            & \multicolumn{4}{c}{Force MAE [eV/\AA{}] $\downarrow$} \\
            Median baseline  & $0.0808$ & $0.0801$ & $0.0787$ & $0.0978$ \\
            CGCNN~\cite{xie2018crystal} & $0.0683$ & $0.0728$ & $0.0670$ & $0.0851$ \\
            SchNet~\cite{schutt2017schnet} & $0.0493$ & $0.0529$ & $0.0509$ & $0.0655$ \\
            SchNet~\cite{schutt2017schnet} -- force-only & $0.0442$ & $0.0469$ & $0.0459$ & $0.0591$ \\
            SchNet~\cite{schutt2017schnet} -- energy-only & $0.5794$ & $0.5974$ & $0.5852$ & $0.6463$ \\
            DimeNet$++$~\cite{klicpera2020directional,klicpera2020fast} & $0.0442$ & $0.0458$ & $0.0444$ & $0.0559$ \\
            DimeNet$++$~\cite{klicpera2020directional,klicpera2020fast} -- force-only & $0.0331$ & $0.0341$ & $0.0340$ & $0.0417$ \\
            DimeNet$++$~\cite{klicpera2020directional,klicpera2020fast} -- energy-only & $0.3399$ & $0.3395$ & $0.3395$ & $0.3643$ \\
            DimeNet$++$~\cite{klicpera2020directional,klicpera2020fast}-Large -- force-only & $0.0280$ & $0.0289$ & $0.0312$ & $0.0371$ \\
            \midrule
            & \multicolumn{4}{c}{Force cosine $\uparrow$} \\
            Median baseline  & $0.0000$ & $0.0000$ & $0.0000$ & $0.0000$ \\
            CGCNN~\cite{xie2018crystal} & $0.1541$ & $0.1369$ & $0.1492$ & $0.1444$ \\
            SchNet~\cite{schutt2017schnet} & $0.3184$ & $0.2954$ & $0.2956$ & $0.2987$ \\
            SchNet~\cite{schutt2017schnet} -- force-only & $0.3595$ & $0.3391$ & $0.3279$ & $0.3403$ \\
            SchNet~\cite{schutt2017schnet} -- energy-only & $0.0845$ & $0.0798$ & $0.0804$ & $0.0830$ \\
            DimeNet$++$~\cite{klicpera2020directional,klicpera2020fast} & $0.3628$ & $0.3476$ & $0.3465$ & $0.3684$ \\
            DimeNet$++$~\cite{klicpera2020directional,klicpera2020fast} -- force-only & $0.4870$ & $0.4717$ & $0.4607$ & $0.4954$ \\
            DimeNet$++$~\cite{klicpera2020directional,klicpera2020fast} -- energy-only & $0.1066$ & $0.0959$ & $0.1048$ & $0.1015$ \\
            DimeNet$++$~\cite{klicpera2020directional,klicpera2020fast}-Large -- force-only & $0.5638$ & $0.5502$ & $0.5115$ & $0.5516$ \\
            \midrule
            & \multicolumn{4}{c}{EFwT $\uparrow$} \\
            Median baseline  & $0.00\%$ & $0.00\%$ & $0.00\%$ & $0.00\%$ \\
            CGCNN~\cite{xie2018crystal} & $0.01\%$ & $0.00\%$ & $0.01\%$ & $0.00\%$ \\
            SchNet~\cite{schutt2017schnet} & $0.11\%$ & $0.04\%$ & $0.06\%$ & $0.01\%$ \\
            SchNet~\cite{schutt2017schnet} -- force-only & $0.00\%$ & $0.00\%$ & $0.00\%$ & $0.00\%$ \\
            SchNet~\cite{schutt2017schnet} -- energy-only & $0.00\%$ & $0.00\%$ & $0.00\%$ & $0.00\%$ \\
            DimeNet$++$~\cite{klicpera2020directional,klicpera2020fast} & $0.10\%$ & $0.03\%$ & $0.05\%$ & $0.01\%$ \\
            DimeNet$++$~\cite{klicpera2020directional,klicpera2020fast} -- force-only & $0.00\%$ & $0.00\%$ & $0.00\%$ & $0.00\%$ \\
            DimeNet$++$~\cite{klicpera2020directional,klicpera2020fast} -- energy-only & $0.00\%$ & $0.00\%$ & $0.00\%$ & $0.00\%$ \\
            DimeNet$++$~\cite{klicpera2020directional,klicpera2020fast}-Large -- force-only & $0.00\%$ & $0.00\%$ & $0.00\%$ & $0.00\%$ \\
         \bottomrule
            \end{tabular}}
    \end{center}
    \caption{Predicting energy and forces from a structure (\gls{S2EF}) as evaluated by Mean Absolute Error (MAE) of the energies, forces MAE, and the percentage of Energies and Forces within Threshold (EFwT). Results reported for models training on the entire training dataset.}
    \label{tab:final_S2EF_subsplit_results}
\end{table}
\setlength{\tabcolsep}{1.4pt}

\begin{figure}[h]
    \centering
    \includegraphics[width=0.8\columnwidth]{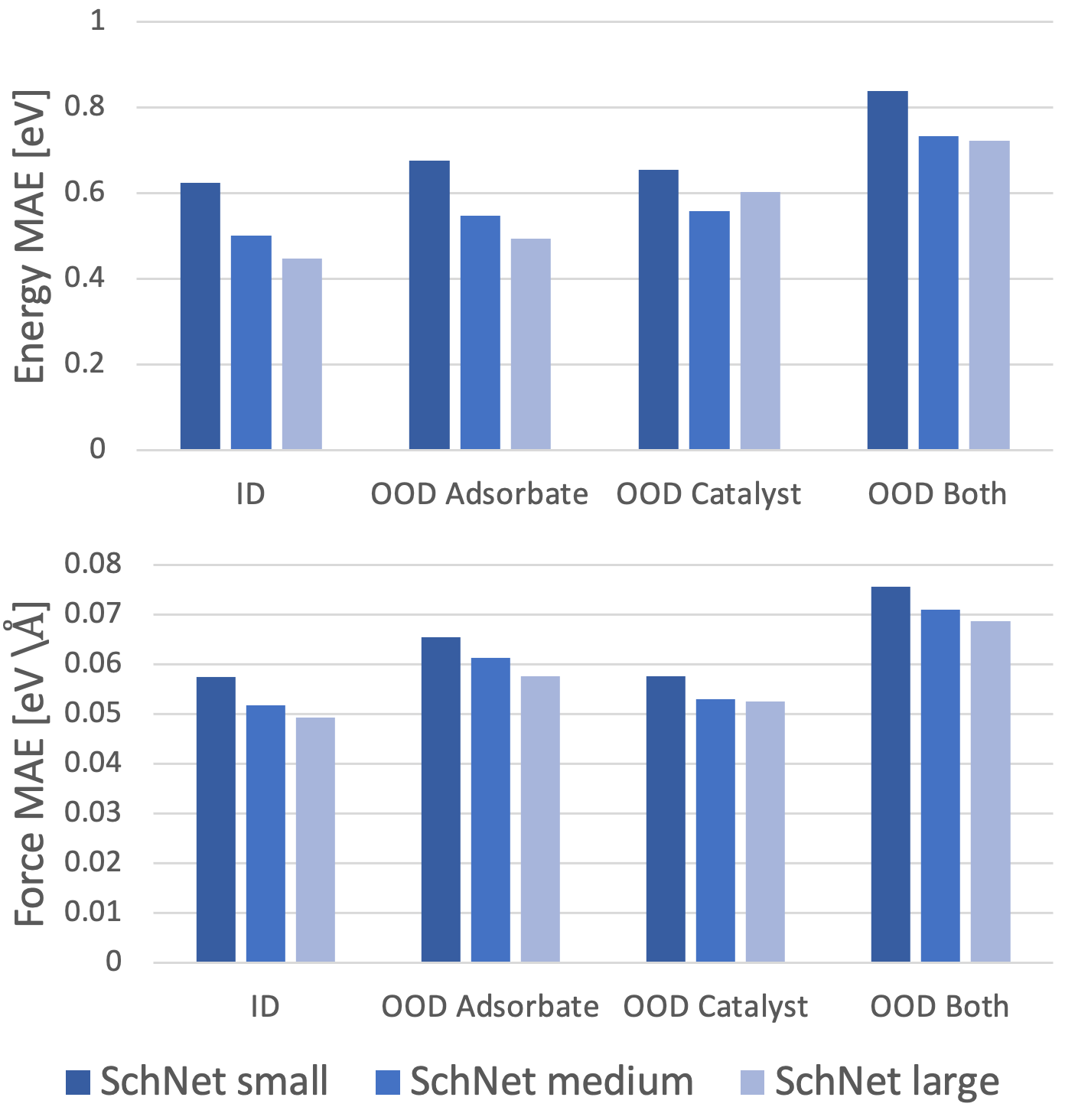}
    \caption{Predicting Structure to Energy and Forces (\gls{S2EF}) as evaluated by Mean Absolute Error (MAE) of the energies and forces.  The small, medium and large SchNet models have the following sizes: Small: 256 hidden, 4 message-passing layers, 1,316,097 params, Medium: 1024 hidden, 3 message-passing layers, 5,704,193 params, Large: 1024 hidden, 4 message-passing layers, 7,396,353 params. Results reported for models trained on the entire training dataset.
    }
    \label{fig:S2EF_schnet_size}
\end{figure}

\setlength{\tabcolsep}{3pt}
\begin{table}[h]
    \begin{center}
        \resizebox{\columnwidth}{!}{
            \begin{tabular}{l cccc }
            & \multicolumn{4}{c}{\gls{IS2RS} Test}  \\
            \midrule
            Model & ID &  OOD Ads & OOD Cat & OOD Both \\
            \midrule
            & \multicolumn{4}{c}{ADwT $\uparrow$} \\
            IS baseline & $21.37\%$ & $19.09\%$ & $21.42\%$ & $26.28\%$\\
            SchNet~\cite{schutt2017schnet} & $15.92\%$ & $12.83\%$ & $14.63\%$ & $14.78\%$\\
            SchNet~\cite{schutt2017schnet} -- force-only & $32.47\%$ & $28.59\%$ & $30.94\%$ & $35.09\%$\\
            DimeNet$++$~\cite{klicpera2020directional,klicpera2020fast} & $30.62\%$ & $26.66\%$ & $30.01\%$ & $32.29\%$ \\
            DimeNet$++$~\cite{klicpera2020directional,klicpera2020fast} -- force-only & $48.73\%$ & $45.19\%$ & $48.54\%$ & $53.17\%$ \\
            DimeNet$++$~\cite{klicpera2020directional,klicpera2020fast}-Large -- force-only & $52.43\%$ & $48.47\%$ & $50.91\%$ & $54.85\%$ \\
            \hline
            & \multicolumn{4}{c}{FbT $\uparrow$} \\
            IS baseline & $0.00\%$ & $0.00\%$ & $0.00\%$ & $0.00\%$\\
            SchNet~\cite{schutt2017schnet} & - & - & - & -  \\
            SchNet~\cite{schutt2017schnet} -- force-only & $0.00\%$ & $0.00\%$ & $0.00\%$ & $0.00\%$\\
            DimeNet$++$~\cite{klicpera2020directional,klicpera2020fast} & $0.00\%$ & $0.20\%$ & $0.00\%$ & $0.00\%$ \\
            DimeNet$++$~\cite{klicpera2020directional,klicpera2020fast} -- force-only & $0.61\%$ & $0.20\%$ & $0.00\%$ & $0.20\%$ \\
            DimeNet$++$~\cite{klicpera2020directional,klicpera2020fast}-Large -- force-only & $1.02\%$ & $0.40\%$ & $0.00\%$ & $0.20\%$ \\
            \hline
            & \multicolumn{4}{c}{AFbT $\uparrow$} \\
            IS baseline & $0.06\%$ & $0.34\%$ & $0.21\%$ & $0.00\%$\\
            SchNet~\cite{schutt2017schnet} & - & - & - & -   \\
            SchNet~\cite{schutt2017schnet} -- force-only & $5.31\%$ & $2.82\%$ & $2.66\%$ & $2.73\%$\\
            DimeNet$++$~\cite{klicpera2020directional,klicpera2020fast} & $3.60\%$ & $3.01\%$ & $2.61\%$ & $2.33\%$ \\
            DimeNet$++$~\cite{klicpera2020directional,klicpera2020fast} -- force-only & $17.42\%$ & $14.67\%$ & $14.12\%$ & $14.46\%$ \\
            DimeNet$++$~\cite{klicpera2020directional,klicpera2020fast}-Large -- force-only & $25.58\%$ & $20.73\%$ & $20.05\%$ & $20.62\%$ \\
            \bottomrule
            \end{tabular}}
    \end{center}
    \caption{Predicting relaxed structure from initial structure (\gls{IS2RS}) as evaluated by Average Distance within Threshold (ADwT), Forces below Threshold (FbT), and Average Forces below Threshold (AFbT). All values in percentages, higher is better. Results reported for structure to force models trained on the All training dataset. The initial structure was used as a naive baseline (IS baseline). FbT and AFbT metrics are only computed when ADwT metrics are greater than 20.26\%.}
    \label{tab:final_structure_results}
\end{table}
\setlength{\tabcolsep}{1.4pt}

\setlength{\tabcolsep}{3pt}
\begin{table*}[t]
    \begin{center}
        \resizebox{\textwidth}{!}{
            \begin{tabular}{l l cccc | cccc  }
            & & \multicolumn{8}{c}{\gls{IS2RE} Test}  \\
            \midrule
            & & \multicolumn{4}{c|}{Energy MAE [eV] $\downarrow$} & \multicolumn{4}{c}{EwT $\uparrow$}  \\
            \cmidrule(l{4pt}r{4pt}){3-6}
            \cmidrule(l{4pt}r{4pt}){7-10}
        Model & Approach & ID &  OOD Ads & OOD Cat & OOD Both & ID &  OOD Ads & OOD Cat & OOD Both \\
            \midrule
                Median baseline & -
                    & $1.7489$ & $1.8911$ & $1.7107$ & $1.6807$
                    & $0.75\%$ & $0.69\%$ & $0.83\%$ & $0.78\%$ \\
                \midrule
                CGCNN~\cite{xie2018crystal} & Direct
                    & $0.6135$ & $0.9155$ & $0.6211$ & $0.8506$
                    & $3.41\%$ & $1.93\%$ & $3.11\%$ & $1.99\%$  \\
                SchNet~\cite{schutt2017schnet} & Direct
                    & $0.6372$ & $0.7342$ & $0.6611$ & $0.7035$
                    & $2.96\%$ & $2.33\%$ & $2.95\%$ & $2.22\%$  \\
                DimeNet$++$~\cite{klicpera2020directional,klicpera2020fast} & Direct
                    & $0.5605$ & $0.7252$ & $0.5750$ & $0.6613$
                    & $4.26\%$ & $2.06\%$ & $4.10\%$ & $2.42\%$ \\
                \midrule
                SchNet~\cite{schutt2017schnet} & Relaxation
                    & $0.7088$ & $0.7741$ & $0.7665$ & $0.8055$
                    & $4.23\%$ & $2.63\%$ & $3.52\%$ & $2.52\%$ \\
                SchNet~\cite{schutt2017schnet} -- force-only + energy-only & Relaxation
                    & $0.7066$ & $0.7420$ & $0.7966$ & $0.7493$
                    & $4.18\%$ & $2.98\%$ & $3.39\%$ & $2.70\%$ \\
                DimeNet${+}{+}$~\cite{klicpera2020directional,klicpera2020fast} & Relaxation
                    & $0.6687$ & $0.6864$ & $0.6858$ & $0.6835$
                    & $4.29\%$ & $3.36\%$ & $3.79\%$ & $3.51\%$ \\
                DimeNet${+}{+}$~\cite{klicpera2020directional,klicpera2020fast} -- force-only + energy-only & Relaxation
                    & $0.5112$ & $0.5744$ & $0.5922$ & $0.6130$
                    & $6.14\%$ & $4.29\%$ & $5.10\%$ & $3.84\%$ \\
                DimeNet${+}{+}$~\cite{klicpera2020directional,klicpera2020fast} -- large force-only + energy-only & Relaxation
                    & $0.5022$ & $0.5430$ & $0.5780$ & $0.6117$
                    & $6.58\%$ & $4.34\%$ & $5.09\%$ & $3.93\%$ \\
            \bottomrule
            \end{tabular}}
    \end{center}
    \caption{Predicting relaxed state energy from initial structure (\gls{IS2RE}) as evaluated by Mean Absolute Error (MAE) of the energies and the percentage of Energies within a Threshold (EwT) of the ground truth energy. Results reported for models trained on the All training dataset.}
    \label{tab:final_energy_subsplit_results}
\end{table*}
\setlength{\tabcolsep}{1.4pt}

\begin{figure*}[h!]
    \centering
    \begin{subfigure}{.3\textwidth}
        \centering
        \includegraphics[width=\linewidth]{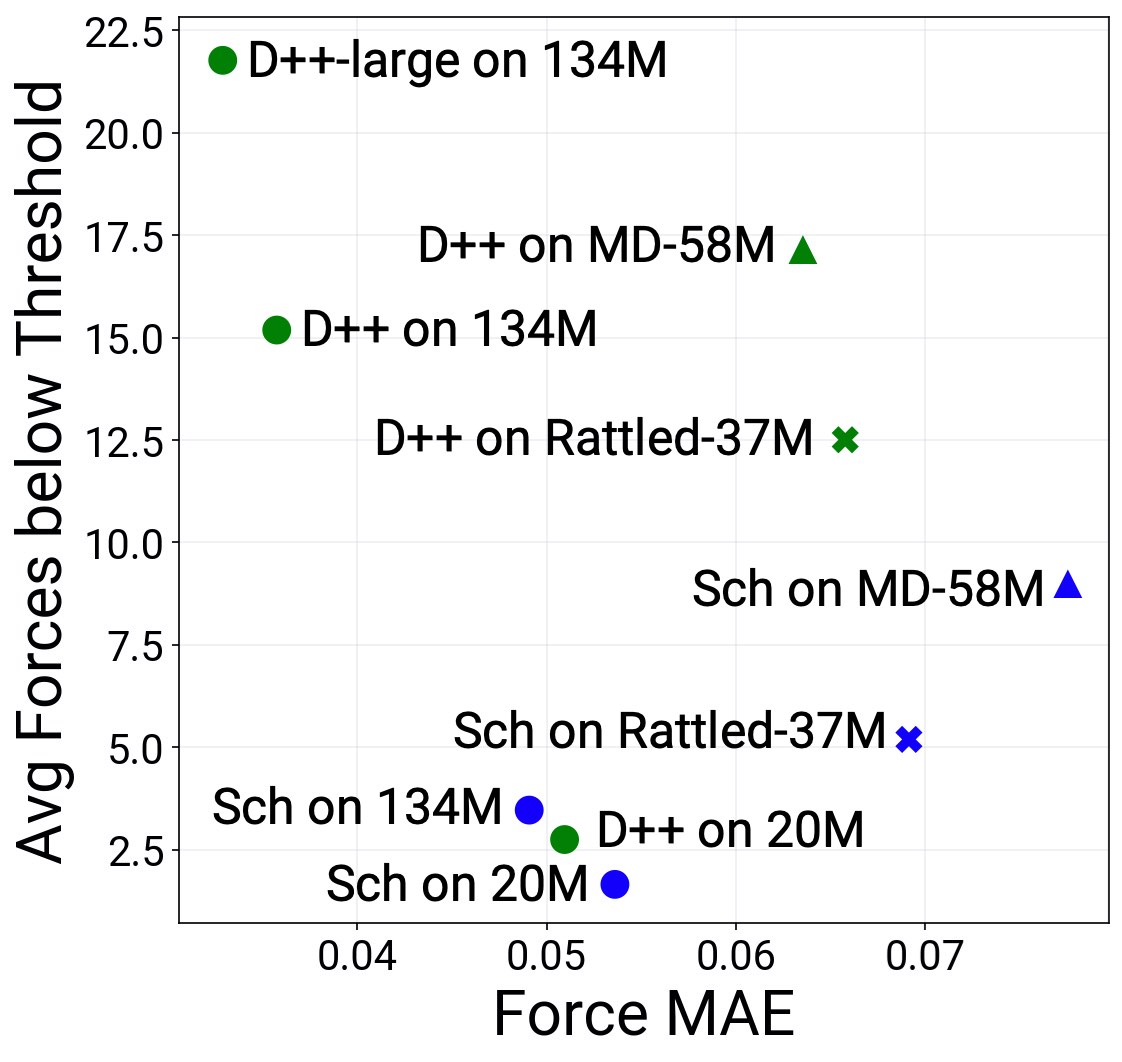}
        \caption{}
        \label{fig:afbt_mae}
    \end{subfigure}
    \quad
    \begin{subfigure}{.3\textwidth}
      \centering
      \includegraphics[width=\linewidth]{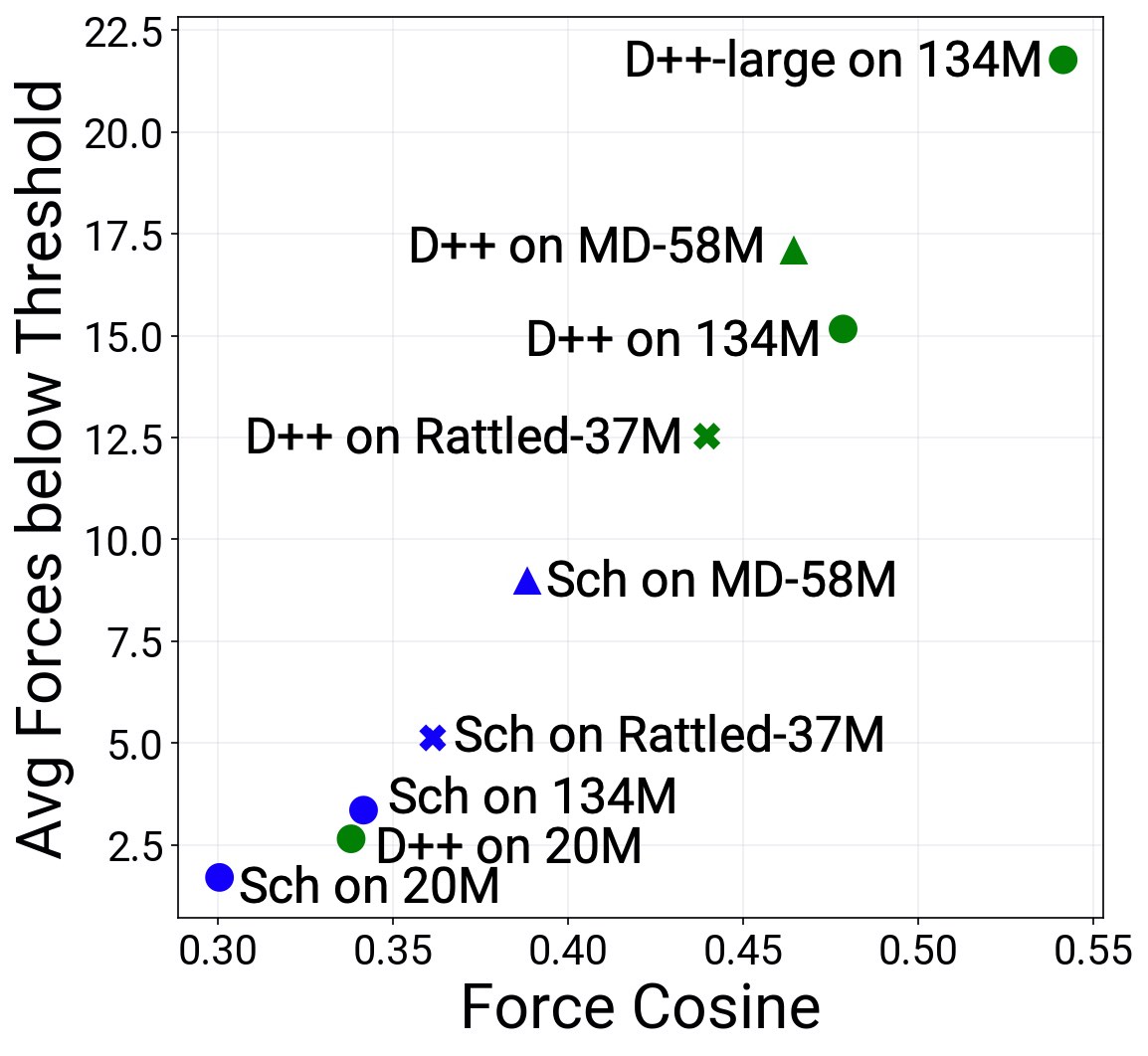}
      \caption{}
      \label{fig:afbt_cosine}
    \end{subfigure}%
    \quad
    \begin{subfigure}{.3\textwidth}
      \centering
      \includegraphics[width=\linewidth]{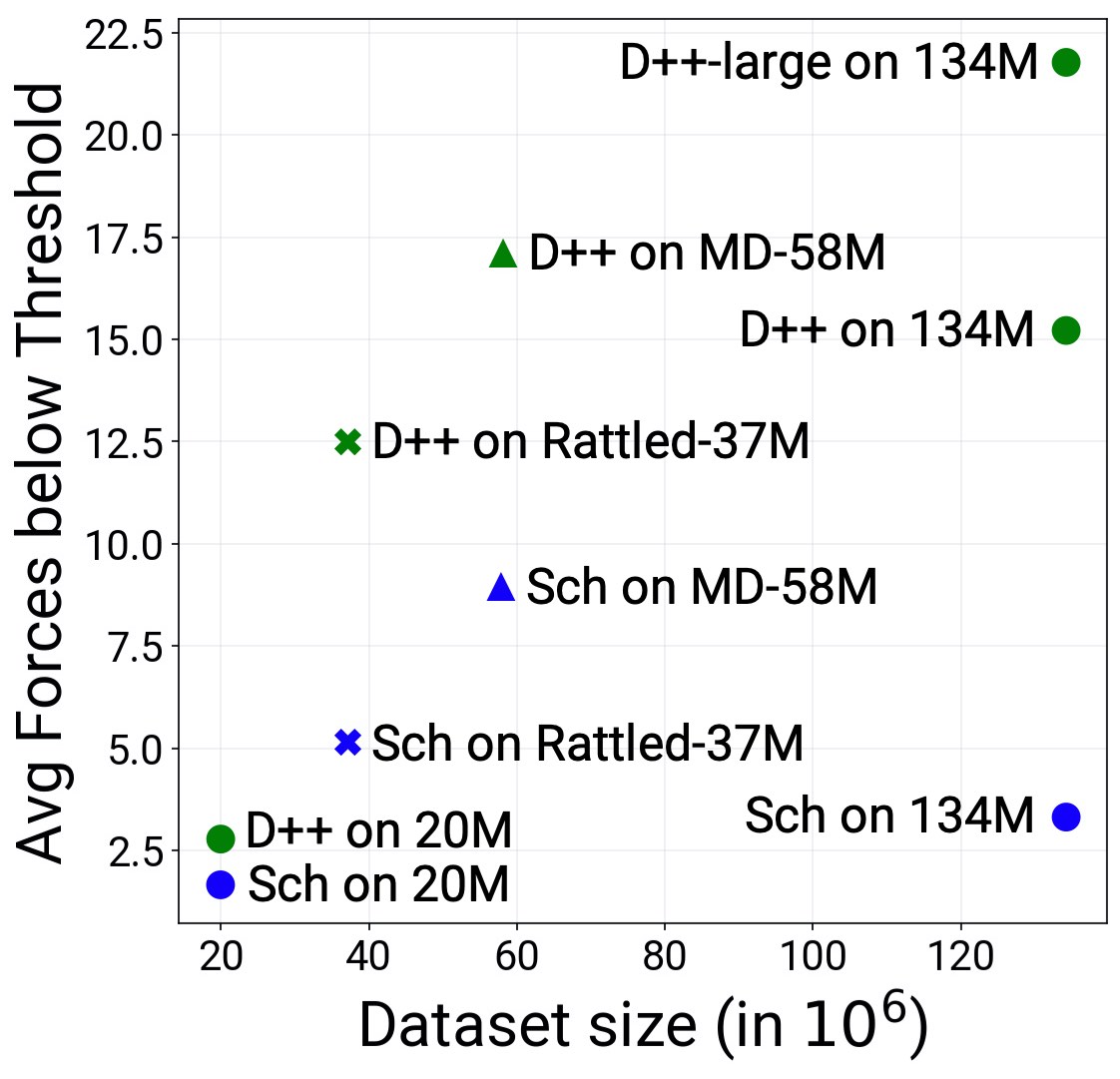}
      \caption{}
      \label{fig:afbt_data}
    \end{subfigure}
    \caption{Results of force-only SchNet (denoted by `Sch') and DimeNet$++$ (`D++')
        \gls{S2EF} models trained on S2EF-$20M$, S2EF-$100M$, S2EF-$20M$ $+$ Rattled (`Rattled-$37M$')
        and S2EF-$20M$ $+$ MD (`MD-$58M$') dataset splits used to
        drive relaxations from given initial structures (\gls{IS2RS}).
        We plot \gls{IS2RS} AFbT performance against \gls{S2EF} force cosine,
        \gls{S2EF} force MAE and number of training samples for the different variants.
        \ref{fig:afbt_mae},\ref{fig:afbt_cosine}: \gls{IS2RS} AFbT seems to correlate
        better with \gls{S2EF} force cosine than \gls{S2EF} force MAE,
        especially when analyzing models trained on Rattled-$37M$ or MD-$58M$ data.
        \ref{fig:afbt_data}: Further, both DimeNet$++$ and SchNet achieve higher
        AFbT when trained on MD-$58M$ than S2EF-$134M$. Additional MD data seems
        to offer a stronger learning signal than additional S2EF data.}
    \label{fig:md_rattled_plots}
\end{figure*}

{\bf \gls{IS2RE}:} Similar to the \gls{S2EF} task we propose two metrics for \gls{IS2RE}. The first measures the MAE on the computed and ground truth energy. The second measures the energies within a threshold (EwT) of the ground truth, which once again measures the percentage of estimated energies that are likely to be practically useful.

\begin{itemize}[leftmargin=*]
    \item[] {\bf Energy MAE:} Mean Absolute Error between the computed relaxed energy and the ground truth relaxed energy.
    \item[] {\bf EwT:} The percentage of computed relaxed energies within $\epsilon = 0.02$ eV of the ground truth relaxed energy.
\end{itemize}

While our evaluation metrics focus on accuracy, it is important to note that methods should also be significantly faster than conventional DFT. As a rough benchmark, we desire energy and force estimates at approximately 10 ms which would significantly improve the applicability of DFT. Significantly faster than this  (closer in speed to classical force fields) would open up even more interesting applications. We ask that users self-report timing results, but we are not going to make that a core part of the challenge as computation time can likely be further optimized for the best models and with hardware acceleration.

\subsection{Leaderboard}
To ensure consistent and fair evaluation, a public leaderboard is available on the Open Catalyst Project webpage (\ocpurl{}). Results on any of the tasks' test datasets may be uploaded for evaluation. Ground truth test data is not publicly released to reduce potential overfitting. Evaluation on the test set may only be done through the leaderboard. Ablation studies and hyper-parameter tuning may be done and reported on using the validation datasets.


\subsection{Results}
To provide baselines for the \gls{OC20} dataset, we report results using three state-of-the-art approaches: CGCNN~\cite{xie2018crystal}, SchNet~\cite{schutt2017schnet}, and DimeNet$++$~\cite{klicpera2020fast, klicpera2020directional}. Details of the models' implementations can be found in the Baselines Section.

\begin{figure*}[h]
    \centering
    \includegraphics[width=\textwidth]{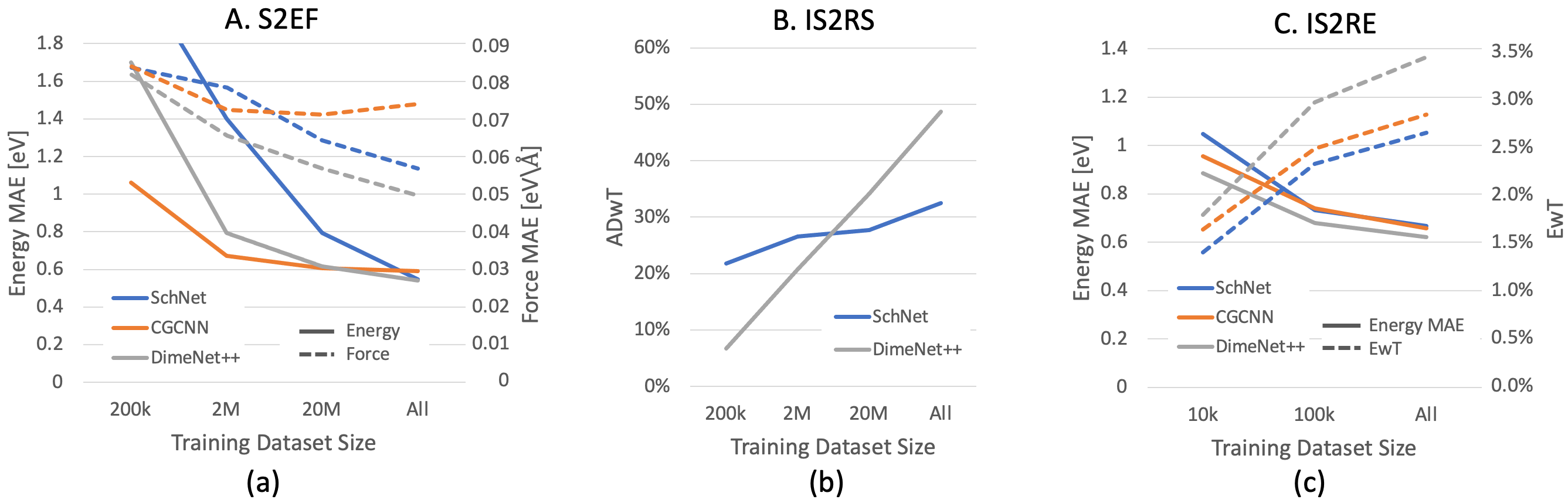}
    \caption{(A) Predicting energy and forces from a structure (\gls{S2EF}) as evaluated by Mean Absolute Error (MAE) of the energies and forces. (B) Predicting relaxed structure from initial structure (\gls{IS2RS}) as evaluated by Average Distance within Threshold (ADwT) using force-only models. (C) Predicting relaxed state energy from initial structure (\gls{IS2RE}) as evaluated by Mean Absolute Error (MAE) of the energies and the percentage of Energies within a Threshold (EwT, $\epsilon = 0.02$ eV) of the ground truth energy. Results reported for \gls{S2EF} and \gls{IS2RS} trained on 200k, 2M, 20M and All dataset sizes. Results reported for \gls{IS2RE} trained on 10k, 100k, and All dataset sizes. \gls{S2EF} and \gls{IS2RE} values averaged across validation subsplits. \gls{IS2RS} values evaluated on the test in-domain (ID) subsplit.}
    \label{fig:dataset_size}
\end{figure*}

{\bf \gls{S2EF}:} Results on CGCNN~\cite{xie2018crystal}, SchNet~\cite{schutt2017schnet}, and DimeNet$++$~\cite{klicpera2020directional, klicpera2020fast} are evaluated. All approaches predict structure energies in their forward pass and per-atom forces by the negative gradient of the predicted
energy with respect to atomic positions~\cite{pukrittayakamee2009simultaneous}. Across most metrics DimeNet$++$ performs the best, with SchNet marginally outperforming DimeNet$++$ and CGCNN on EFwT. SchNet outperforms CGCNN across all metrics. Since tradeoffs exist in the prediction of energy and forces, we trained three variants of SchNet and DimeNet$++$ with $\{\lambda_E, \lambda_F\} = \{1, 30\}, \{0, 100\}, \{100, 1\}$ for SchNet/DimeNet$++$, SchNet/DimeNet$++$ force-only and SchNet/DimeNet$++$ energy-only respectively. As expected, the energy-only model performs best on energy MAE, while the force-only performs best on force MAE. DimeNet$++$ and SchNet both provide a balance between the two and the best results on EFwT. All approaches perform badly on the EFwT metric; indicating that the results are still far from being practically useful. Table \ref{tab:final_S2EF_subsplit_results} and Figure \ref{fig:S2EF_schnet_size} show results across subsplits. As expected, the In Domain (ID) achieves the best results and the OOD Both performs the worst. However, results are not dramatically different between In Domain, OOD Adsorbate and OOD Catalyst, which shows some generalization to new adsorbates and catalysts. Increases in training data sizes results in significant improvements, Figure \ref{fig:dataset_size}(A). The rate and amount of improvement varies based on the model. Finally, wider and deeper models are shown to improve accuracies in Figure \ref{fig:S2EF_schnet_size}. Both increased depth (Medium to Large) and width (Small to Medium) show improvements.

{\bf \gls{IS2RS}:}
For \gls{IS2RS}, we use our \gls{S2EF} baselines to drive ML relaxations from the given initial structures to estimate the relaxed structures using L-BGFS~\cite{liu1989limited}, examples are shown in Figure \ref{fig:relaxations}(A). Table \ref{tab:final_structure_results} shows that DimeNet$++$ outperforms SchNet in the ADwT and AFbT metrics. However, the FbT metrics indicate both methods do not produce relaxed structures with forces below thresholds used in practice. Since only the computed forces are used for the IS2RS task and not the energies, it is not surprising that the DimeNet$++$ force-only model performs the best. It was trained using only force losses and performs significantly better on AFbT and ADwT, but still is near zero when measured by FbT. A plot of FbT across thresholds from 0.01 to 0.6 for SchNet is shown in Figure \ref{fig:relaxations}(C). Both methods show better generalization to new adsorbates vs new catalyst material compositions. Similar to \gls{S2EF} improved results are found with more training data, especially for DimeNet$++$ and SchNet, Figure \ref{fig:dataset_size}(B). Experiments using the additional rattled and MD data are shown in Figure \ref{fig:md_rattled_plots}. Interestingly, the force cosine metric appears to better correlate with AFbT scores than force MAE. A discussion on these results may be found in the supplementary.

{\bf \gls{IS2RE}:}
For \gls{IS2RE} we explore two pathways for computing the relaxed energy from the initial state, Figure \ref{fig:relaxations}(B). The first directly computes the relaxed energy given the initial state. The same model architectures are used as the \gls{S2EF} task, but with new weights learned. The second approach uses models trained on the \gls{S2EF} task to perform ML relaxations from which the resulting energy is returned. Note that the ML relaxation approach is about 200 times more expensive to compute, since energies needs to be computed at each relaxation step.


As shown in Table \ref{tab:final_energy_subsplit_results}, the hybrid relaxation approaches outperformed the direct across all metrics. The percentage of predicted energies within a tight threshold (EwT) ranged from $2\%$ to $6\%$; indicating that accuracies are still below practical usefulness. Generalization to new catalyst compositions performed better than new adsorbates. As shown in Figure \ref{fig:dataset_size}(C), larger dataset sizes could significantly improve performance. The best direct-based approach, DimeNet$++$, was evaluated via the relaxation-based approach.  The use of DimeNet$++$ force-only to perform the relaxation, followed by DimeNet$++$ energy-only to compute the relaxed energy significantly outperformed the use of a single model (optimized for EFwT) to compute both. Best metrics were achieved using the large DimeNet$++$ force-only model, followed by DimeNet$++$ energy-only.

\begin{figure*}[ht!]
    \centering
    \includegraphics[width=0.78\textwidth]{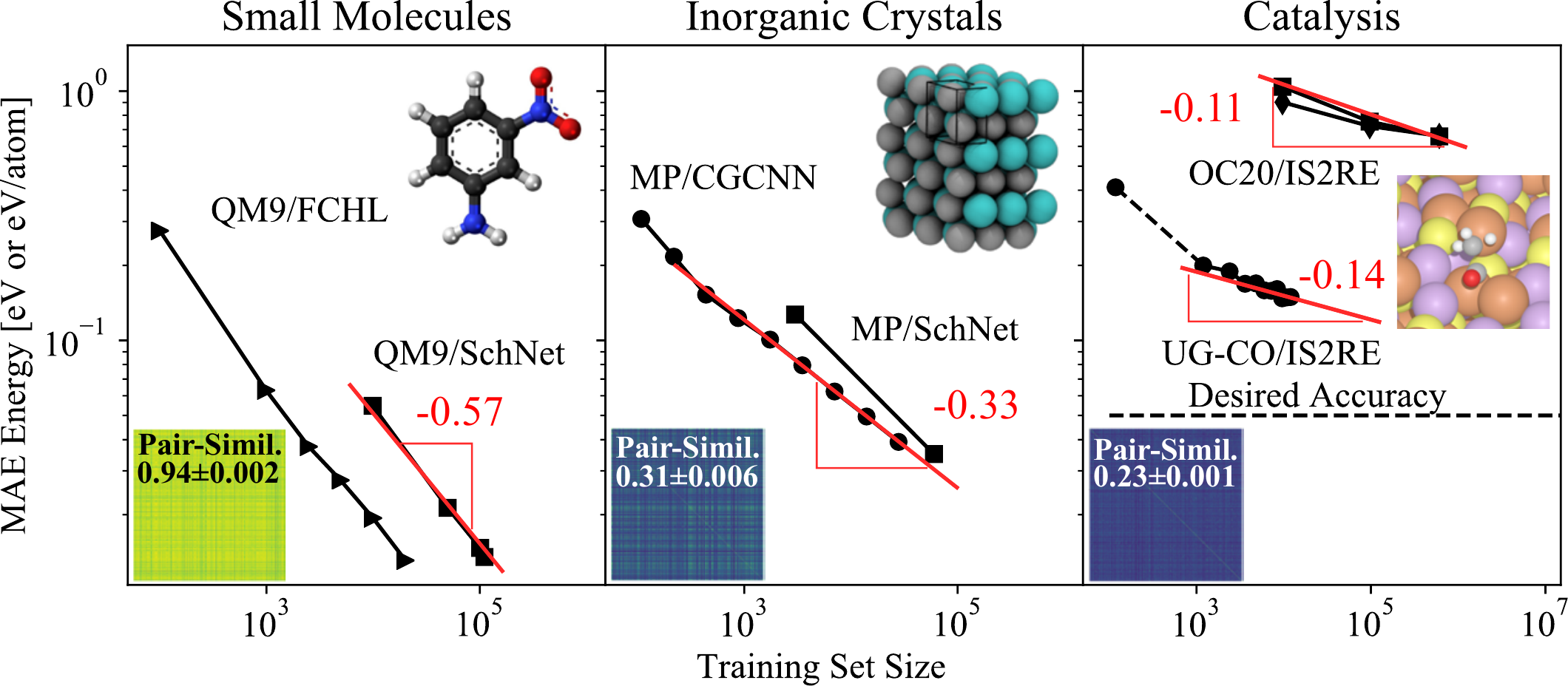}
    \caption{Model performance versus dataset size across three related atomistic domains. Insets are pairwise similarity for selected structures from the respective dataset using GraphDot (see the SI for details) (0/dark-blue/not-similar to 1/yellow/identical)\cite{GraphdotTang2020, Tang2019}. (left) Results \cite{von2020exploring} for FCHL/SchNet models trained on the QM9 small molecule dataset (slope -0.57). (middle) Models\cite{xie2018crystal,schutt2017schnet} trained on Materials Project formation energies (slope -0.33, more difficult). (right) Results for catalysis including a literature dataset for CO adsorbates \cite{Tran2018} and this work (slope -0.11 to -0.14, most difficult). Note that reaching the desired accuracy will require several orders of magnitude more data with current models.
    \label{fig:scaling}}
\end{figure*}

\section{Outlook and Future Directions}

The baseline models in this work give significant insights into the complexity of day-to-day challenges in catalysis and what it will take to achieve generalizable models. Motivated by previous efforts\cite{huang2018fundamentals}, we analyzed model performance for increasing dataset sizes to illustrate the differences between catalysis and related efforts---e.g., materials sciences or small molecule property prediction. Figure \ref{fig:scaling}(left) and Figure \ref{fig:scaling}(middle) show the performance of GNN models similar to the baseline models in this work on datasets for small molecules (QM9) and materials (formation energies from the materials project). The scaling of model accuracy with respect to dataset size is related to the effective dimensionality of the task and the effective representation in the model. Comparing DimeNet$++$ performance across all three tasks shows that the aggressive scaling for small molecules is reduced for inorganic materials, and further reduced for surfaces. Focusing on results from this study in Figure \ref{fig:scaling}(right) shows that the scaling is similar for the same baseline models trained on the OC20 dataset and a related literature dataset of CO adsorption energies (see the SI). Importantly, this suggests that achieving the desired accuracy using the current baseline models would require a dataset nearly 10 orders of magnitude larger than the current dataset. This implies that this problem will not be solved through brute-force methods alone, and that significantly improved ML representations are also necessary.  This is an exciting opportunity for the broader community.

For the computer science and ML communities, we expect that this dataset will provide unique challenges and spur innovation in atomistic simulations. Many state-of-the-art methods for organic and inorganic materials are based on graph convolutional networks~\cite{xie2018crystal}, which have seen rapid progress. With the above perspective, we expect that additional creative solutions will be necessary to fully solve these tasks. While they have not been demonstrated for inorganic materials, physics-informed tensor representations for small molecules may be helpful~\cite{miller2020relevance, bratholm2020community, anderson2019cormorant, nigam2020recursive}. Element embeddings and representations will be important to scale across materials. Incorporation of lower-level physics-based potentials is welcomed and encouraged. This includes the use of related datasets (organic molecules or inorganic materials) for pre-training or learning priors. Incorporating other electronic features in the training set, such as charge distribution to correctly localize effects is also an opportunity to effectively reduce the dimensionality of the problem.

Note that the size of this dataset is larger by 2 orders of magnitude than previous catalyst DFT dataset efforts \cite{Tran2018, Hummelshoj2012a}. Along with the potential for more accurate ML models, it provides practical challenges to training atomistic machine learning models at scale, similar to software engineering challenges in image recognition and NLP~\cite{he2016deep,radford2019language}. The largest baseline models with \textit{ca.} 10 million parameters were trained on upwards of 32 GPUs at a time, so we encourage the catalysis community to take advantage of these GPU-enabled resources. This is well-timed with the wave of large GPU-enabled supercomputers that are well-suited to these challenges, such as Perlmutter (DOE NERSC) or Summit (DOE OLCF), among many others.

The baseline models in this work represent the state-of-the-art for deep learning methods to predict thermochemistry for small molecules on inorganic surfaces. Solving this challenge with future model development efforts would enable a new generation of computational chemistry methods. In particular, on-the-fly thermochemistry for reaction intermediates would enable reaction mechanism prediction across materials or composition space. Accelerated methods would also enable the more routine use of more accurate computational methods (e.g. hybrid, exact-exchange, or RPA calculations) by focusing these efforts on the most promising and pre-relaxed structures. A solution to the S2EF task would enable transition state calculations, kinetic approximations, vibrational frequency calculations, and the more routine use of long timescale molecular dynamics for studying these systems. Sensitivity analyses will be necessary to understand the level of accuracy needed for models to be practically relevant for varying applications. Given the sparsity and breadth of OC20, the availability of relevant experimental data will also be a crucial challenge in the next stage of validating model results with experiments. The potential applicability of the OC20 dataset is not just catalysis, but also has implications for areas where organic and inorganic materials interact, such as water quality remediation, geochemistry, advanced manufacturing, and durable energy materials.

\begin{suppinfo}

The supporting information contains details on the precise DFT calculation methods, the adsorption energy reference energies, the adsorbates and their assuming binding configurations, details on graph construction, description of the graph similarity metrics, a few sample GFN0-XTB relaxations, the precise train/test/validation splits, details on the modified CGCNN/SchNet/DimeNet$++$ implementations, results on the Rattled/MD experiments, hyperparameters for baseline models, a list of adsorbates in \dataset{}, and full results on the validation splits. The full open dataset is provided at \ocpurl{} in accessible extxyz format, and the baseline models are provided as an open source repository at \baselinesurl{}.

\end{suppinfo}

\begin{acknowledgement}

This research used resources of the National Energy Research Scientific Computing Center (NERSC), a U.S. Department of Energy Office of Science User Facility operated under Contract No. DE-AC02-05CH11231. B.W. acknowledges support from the NERSC Early Science Application Program. The authors acknowledge very helpful discussions with John Kitchin (CMU), Dionisios Vlachos (UD), and Philippe Sautet (UCLA) on the dataset construction and usage.

\end{acknowledgement}

\bibliography{main_arxiv}

\begin{mcitethebibliography}{112}
\bibitem[vas()]{vasp-license}
``The calculations in this work have been performed using the ab-initio
  total-energy and molecular- dynamics package VASP (Vienna ab-initio
  simulation package) developed at the Institut f{\"u}r Materialphysik of the
  Universi{\"a}t Wien2,3.". \relax
\mciteBstWouldAddEndPunctfalse
\mciteSetBstMidEndSepPunct{\mcitedefaultmidpunct}
{}{\mcitedefaultseppunct}\relax
\EndOfBibitem
\end{mcitethebibliography}

\clearpage

\section{DFT Relaxations}

\begin{figure}[t]
    \centering
    \includegraphics[width=0.7\columnwidth]{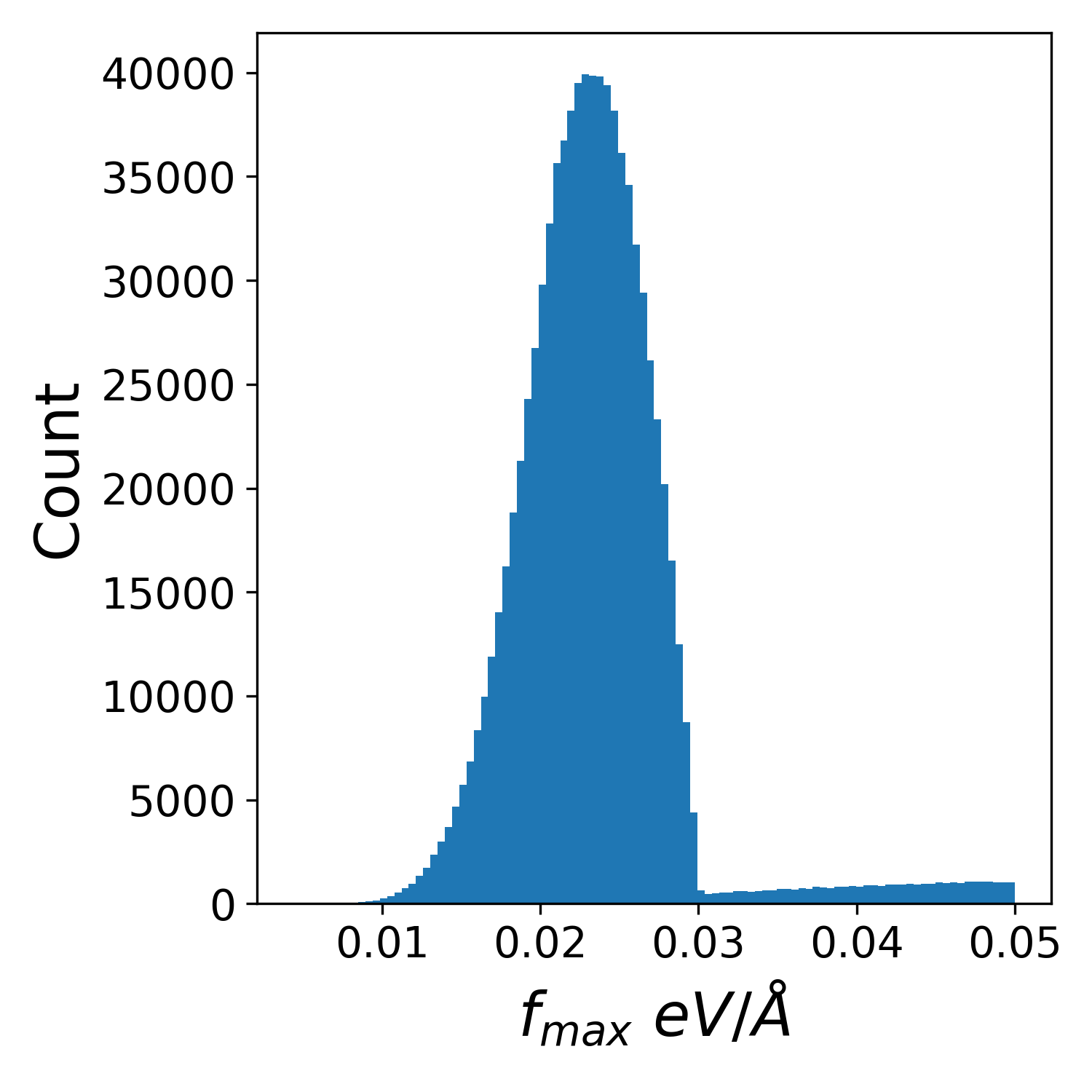}
    \caption{The distribution of max-absolute forces, $f_{max}$, for systems that converged and completed successfully. Systems in which $f_{max} > 0.05$ eV/{\AA} were excluded from all tasks except S2EF.}
    \label{fig:forces}
\end{figure}

DFT calculations were performed with the \textit{Vienna Ab Initio Simulation Package} (VASP)\cite{Kresse1994, Kresse1996, Kresse1996a, vasp-license, kresse1999ultrasoft} with periodic boundary conditions and the projector augmented wave (PAW) pseudopotentials~\cite{blochl1994projector, kresse1999ultrasoft}. The external electrons were expanded in plane waves with kinetic energy cut-offs of 350 eV. Exchange and correlation effects were taken into account via the generalized gradient approximation~\cite{perdew1996generalized} and the revised Perdew-Burke-Ernzerhof (RPBE) functional, because of its improved description of the energetics of atomic and molecular bonding to surfaces~\cite{hammer1999improved}. Bulk and surface calculations were performed considering a K-point mesh for the Brillouin zone derived from the unit cell parameters as an on-the-spot method, employing the Monkhorst-Pack grid~\cite{monkhorst1976special}. The ionic degrees of freedom were relaxed using a Conjugate Gradient minimization~\cite{teter1989solution, press1986numerical}. The relaxation was terminated when either the Hellmann-Feynman forces~\cite{feynman1939forces}  were less than 0.03 eV/\AA{} or the relaxation required more than 200 steps in a single uninterrupted VASP call. This limit was reset each time the calculation was checkpointed allowing some relaxations to exceed this 200 steps.  The final distribution of residual forces is shown in Figure \ref{fig:forces} in the SI. Relaxations still converging after approximately 5,000 core-hours were terminated and not included in the dataset. For the electronic degrees of freedom, the energy convergence criteria was fixed to $10^{-4}$ eV, where no spin magnetism or dispersion corrections were included.

\section{Adsorption Energy}

\begin{align*}
    E_{ad} = E_{sys} – E_{slab} – E_{gas}
\end{align*}

Gas phase references, $E_{gas}$, for each adsorbate was computed as a linear combination of N$_2$, H$_2$O, CO, and H$_2$ resulting in the atomic energies from Table~\ref{tab:gas_ref}.

\begin{table}[ht]
    \begin{tabular}{c|c}
        \toprule
        Adsorbate atom & Energy (eV) \\
        \midrule
        H & -3.477 \\
        O & -7.204 \\
        C & -7.282 \\
        N & -8.083 \\
        \bottomrule
    \end{tabular}
    \caption{The per atom energy of individual adsorbate atoms used to calculate the gas phase reference energy for an adsorbate molecule}
    \label{tab:gas_ref}
\end{table}

\section{Computational Workflow}

\begin{figure*}
    \centering
    \includegraphics[width=0.9\textwidth]{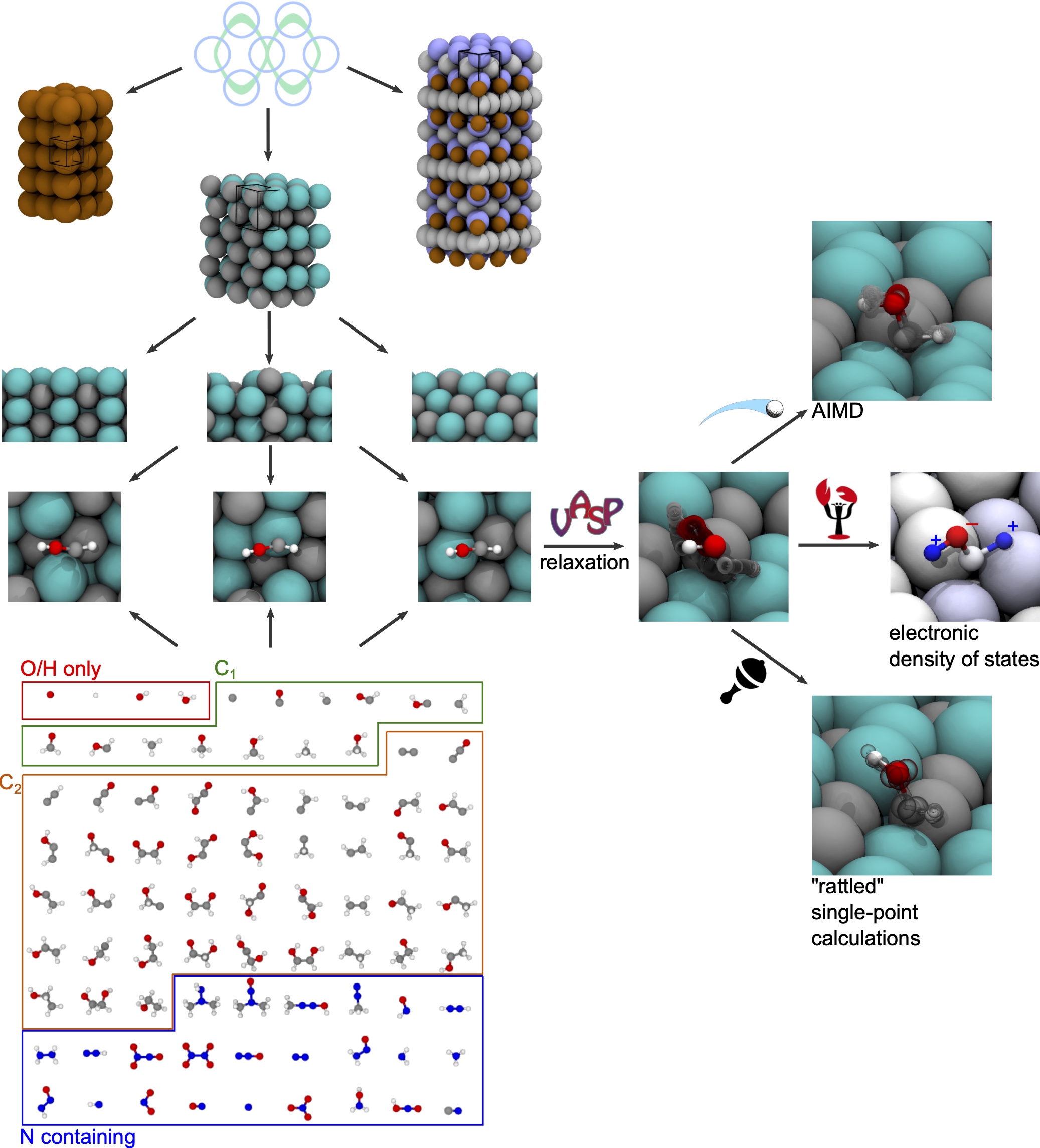}
    \caption{The workflow used to generate the Open Catalyst Dataset.
        Stable materials were downloaded from The Materials Project\cite{Jain2013} and paired with heuristically chosen adsorbates to create adsorption structures.
        These structures were randomly sampled for DFT relaxation and then subsequent AIMD, electronic structure analysis, and single-point rattling calculations.}
    \label{fig:workflow}
\end{figure*}

An illustration of the workflow used to sampled from the dataset and perform calculations is show in Figure \ref{fig:workflow}.

\section{Graph Construction}
\begin{figure}
     \centering
     \includegraphics{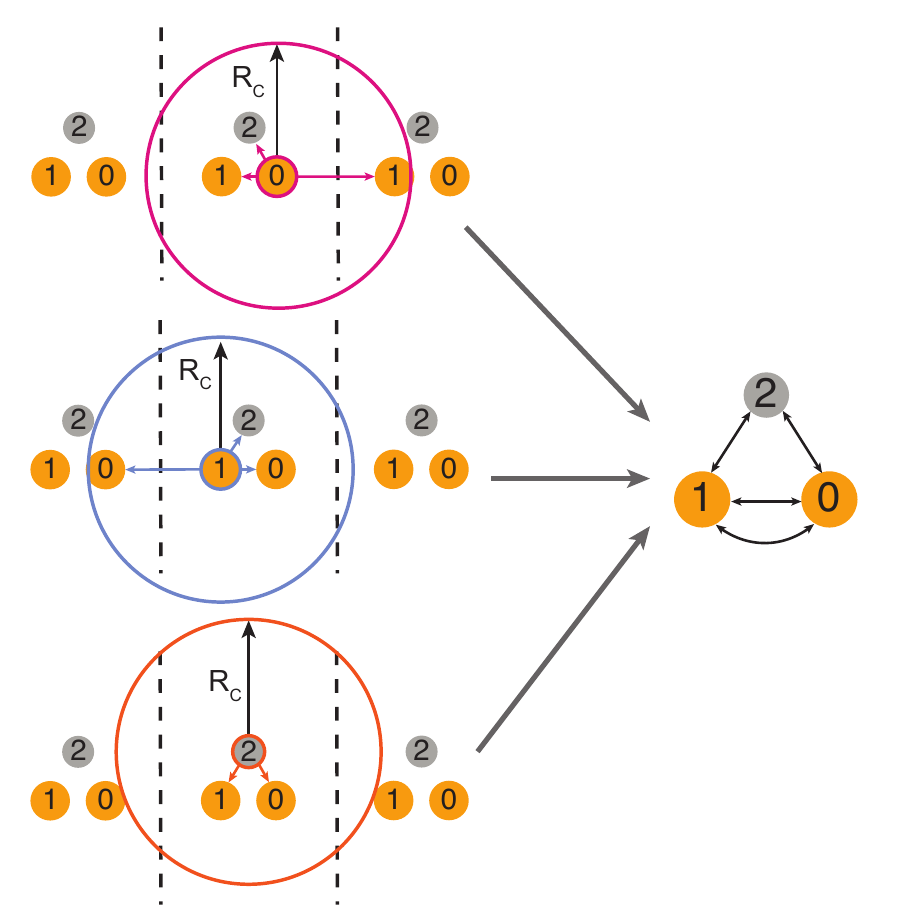}
     \caption{A simple example of constructing a radius graph with periodic boundary conditions. The graph on the right represents all edges assuming each atom as the center node individually (shown on the left).}
     \label{fig:graph_ex}
 \end{figure}

Given a set of atoms in the 3D unit cell that is periodically repeated, we construct a radius graph where nodes represent the atoms and edges represent nearby interaction between pairs of atoms. Specifically, we draw a directed edge from atom $j$ to atom $i$ if atom $j$ is within the cutoff distance from atom $i$, and vice versa. This means that the edges are always bidirectional. Furthermore, since the nodes are periodically repeated, two atoms may have multiple directed edges if they lie within the cutoff distance in multiple repeated cells. If an atom $i$ has more than one edge to an atom $j$, each edge represents atom $j$ in a different cell, resulting in unique relative distances and edge features, Figure \ref{fig:graph_ex}. From the atom-centric view, the above directed multi-graph representation of the atomic system precisely captures the local 3D structure surrounding each atom, taking periodic boundary condition into account.

\section{Graph Pairwise Similarity}
The mean pairwise similarity (mps) between a collection of graphs gives an indication of the diversity present in a given dataset and is comparable between different datasets. Pairwise similarity was computed as the mean of the elements in the upper triangle of the similarity matrix ($\mathbf{K}$) without the diagonal elements included (Equation below). The similarity matrix was calculated using graphs and the molecular kernel from the GraphDot package (https://graphdot.readthedocs.io/en/latest/), details of these methods are provided by Tang et al.~\cite{Tang2019}. Mean pairwise similarity values range from 1, where all graphs are the same and decay to 0. The mean pairwise similarity can be compared between datasets if the graph and the kernel parameters are consistent. For the results in Figure 6 of the main text, we randomly sampled 1000 systems (N) from a 10,000 subsample of each respective dataset and computed the mean pairwise similarity, this was repeated six times to collect statistics. Random subsampling was done to keep the similarity matrix the same size across datasets and to decrease the computational cost. For the similarity matrix calculation the adjacency length scale used to convert atomic structures to graphs was set to 6 \AA{} and the molecular kernel edge length scale was set to 18 \AA{}, nearly identical results were achieved with 2 \AA{} and 5 \AA{} respectively. All other parameters were set to default values.

\begin{align*}
    \text{mps} = \frac{1}{N(N-1)/2} \sum_{i,j}^{N} \mathbf{K}_{ij} \\ \text{where} \ i < j \ \ \ \ \ \
    \label{eq:mps}
\end{align*}

\section{Baseline Models Implementation}
All proposed baseline models were implemented using PyTorch Geometric. Several implementation changes, however, were necessary to make such models relevant to our dataset and tasks. We outline the modifications below:

\subsubsection{SchNet}
\begin{itemize}
    \item Periodic boundary conditions (PBCs) were incorporated into the PyTorch Geometric implementation of SchNet.
\end{itemize}

\subsubsection{DimeNet$++$}
\begin{itemize}
    \item PBCs were incorporated into the PyTorch Geometric implementation of DimeNet$++$.
\end{itemize}

\subsubsection{CGCNN}
\begin{itemize}
    \item Similar to SchNet, a Gaussian basis function was incorporated to the edge features. Although not contained within the original CGCNN implementation, a significant performance increase was observed.
    \item In order to make force predictions, a gradient call was included in the forward pass with respect to positions. The original CGCNN implementation was only concerned with energy predictions.
\end{itemize}

\section{Hyperparameters for Baseline Models}
Model hyperparameters for the `All' splits of the IS2RE and S2EF tasks are provided in Tables
\ref{tab:cgcnnhp}, \ref{tab:schnethp}, and \ref{tab:dimenethp}.
Hyperparameters of the remaining splits can be found in the corresponding repo: \url{https://github.com/Open-Catalyst-Project/ocp/tree/master/configs}.

\setlength{\tabcolsep}{3pt}
\begin{table*}[h]
        \begin{tabular}{c c c }
            \midrule\midrule
            Hyperparameters  & IS2RE & S2EF \\
            \midrule\midrule
            Size of atom embeddings & 384 & 512\\
            Size of fully connected layers & 512 & 128\\
            Number of fully connected layers & 4 & 3\\
            Number of graph convolutional layers & 6 & 3\\
            Number of Gaussians used for smearing & 100 & 100\\
            Cutoff distance for interatomic interactions & 6 & 6\\
            Batch size (per gpu) & 16 & 24\\
            Initial learning rate & 0.001 & 0.0005\\
            Learning rate gamma & 0.1 & 0.1\\
            Learning rate milestones & [5, 9, 13] & [3, 5, 7]\\
            Warmup epochs & 3 & 2\\
            Warmup factor & 0.2 & 0.2\\
            Max epochs & 20 & 20\\
            Force coefficient & N/A & 10\\
            \bottomrule\bottomrule
        \end{tabular}
    \caption{CGCNN~\cite{xie2018crystal} hyperparameters on the All split of the IS2RE and S2EF tasks.}
    \label{tab:cgcnnhp}
\end{table*}
\setlength{\tabcolsep}{1.4pt}

\setlength{\tabcolsep}{3pt}
\begin{table*}[ht]
        \begin{tabular}{c c c }
            \midrule\midrule
            Hyperparameters  & IS2RE & S2EF \\
            \midrule\midrule
            Number of hidden channels & 384 & 1024\\
            Number of filters & 128 & 256\\
            Number of interaction blocks & 4 & 5\\
            Number of Gaussians used for smearing & 100 & 200\\
            Cutoff distance for interatomic interactions & 6 & 6\\
            Global aggregation & add & add\\
            Batch size (per gpu) & 64 & 20\\
            Initial learning rate & 0.001 & 0.0001\\
            Learning rate gamma & 0.1 & 0.1\\
            Learning rate milestones & [10, 15, 20] & [3, 5, 7]\\
            Warmup epochs & 3 & 2\\
            Warmup factor & 0.2 & 0.2\\
            Max epochs & 30 & 15\\
            Force coefficient & N/A & 30\\
            \bottomrule\bottomrule
        \end{tabular}
    \caption{SchNet~\cite{schutt2017schnet} hyperparameters on the All split of the IS2RE and S2EF tasks.}
    \label{tab:schnethp}
\end{table*}
\setlength{\tabcolsep}{1.4pt}

\setlength{\tabcolsep}{3pt}
\begin{table*}[ht]
        \begin{tabular}{c c c }
            \midrule\midrule
            Hyperparameters  & IS2RE & S2EF \\
            \midrule\midrule
            Number of hidden channels & 256 & 192\\
            Output block embedding size & 192 & 192\\
            Number of interaction blocks & 3 & 3\\
            Number of radial basis functions & 6 & 6\\
            Number of spherical harmonics & 7 & 7\\
            Number of residual layers before skip connection & 1 & 1\\
            Number of residual layers after skip connection & 2 & 2\\
            Number of linear layers in output blocks & 3 & 3\\
            Cutoff distance for interatomic interactions & 6 & 6\\
            Batch size (per GPU) & 4 & 8\\
            Initial learning rate & 0.0001 & 0.0001\\
            Learning rate gamma & 0.1 & 0.1\\
            Learning rate milestones & [4, 8, 12] & [2, 3, 4]\\
            Warmup epochs & 2 & 2\\
            Warmup factor & 0.2 & 0.2\\
            Max epochs & 20 & 7\\
            Force coefficient & N/A & 50\\
            \bottomrule\bottomrule
        \end{tabular}
    \caption{DimeNet$++$~\cite{klicpera2020directional, klicpera2020fast} hyperparameters on the All split of the IS2RE and S2EF tasks.}
    \label{tab:dimenethp}
\end{table*}
\setlength{\tabcolsep}{1.4pt}
\clearpage

\section{IS2RE Performance of Baseline Models on Previous Datasets}

The MAE metrics of the baseline models for the \gls{IS2RE} task are significantly higher than have been reported in recent studies applying ML models to predict adsorption energies\cite{back2019convolutional, tran2020methods, gu2020practical}. There are three key differences in this work. First, the dataset here is larger, more diverse, sparser, and more uniformly sampled than previous datasets making this task more challenging. Second, we are using a more difficult definition of the \gls{IS2RE} task - predict the final energy directly from the initial structure, rather than a clean representation of the final structure \cite{back2019convolutional}. Finally, the baseline models themselves are somewhat different (both implementation, and details of the training and precise form).

To test that the baselines models were consistent with previous efforts, we applied all three models to the \gls{IS2RE} task for a literature dataset of CO adsorption energies \cite{back2019convolutional,Tran2018}, show in Table \ref{tab:ulissi_co}. Our results are consistent, and often better, than previously reported validation accuracy for a CGCNN-based model at approximately 0.190 eV MAE on the literature dataset. This is far lower than the 0.57 eV MAE for our baseline models trained only on the CO subset of the OC20 dataset. This suggests that the dataset diversity is the dominant factor in this variation, and further emphasizes that a uniformly sampled dataset can be more difficult to fit than one obtained through an active learning process that emphasizes high-performing catalysts.

\setlength{\tabcolsep}{3pt}
\begin{table*}[ht]
        \begin{tabular}{c c }
            \midrule
            Model  & Validation \\
            \midrule
            & \multicolumn{1}{c}{Energy MAE [eV] $\downarrow$} \\
            Previous Work~\cite{back2019convolutional, Tran2018} & $0.190$\\
            CGCNN~\cite{xie2018crystal} & $0.174$ \\
            SchNet~\cite{schutt2017schnet} & $0.170$ \\
            DimeNet$++$~\cite{klicpera2020directional, klicpera2020fast} & $0.149$  \\
            \bottomrule
        \end{tabular}
    \caption{Benchmark of our baseline models' implementations on a literature CO dataset\cite{back2019convolutional, Tran2018} as evaluated by Energy MAE.}
    \label{tab:ulissi_co}
\end{table*}
\setlength{\tabcolsep}{1.4pt}

\section{Adsorbates Included}

The full list of adsorbates is indicated in Table \ref{tab:adsorbates}. This list was constructed by considering the four monatomic species and adding common intermediates for renewable energy challenges. The number of possible organic molecules is combinatorially large, so this is not a comprehensive list. Larger molecules (e.g. C3) are also relevant but have an even larger number of possible configurations. Most adsorbates were mono-dentate (binding through a single adsorbate atom), but larger molecules known to bind in bi-dentate configurations were initialized that way. The atoms considered for either mono-dentate or bi-dentate adsorption location is indicated by *.

\begin{table*}[t]
    \centering
    \begin{tabular}{|m{3cm}|m{3cm}|m{11cm}|}
        \hline
        Adsorbate class     &   \# of adsorbates &   Adsorbates \\
        \hline & & \\
        \ce{O}$/$\ce{H} Only   &   4   &
                                \ce{^{*}H},
                                \ce{^{*}O},
                                \ce{^{*}OH},
                                \ce{^{*}OH2} \\ & & \\
        \hline & & \\
        C\textsubscript{1}  &   13  &
                                \ce{^{*}C},
                                \ce{^{*}CO},
                                \ce{^{*}CH},
                                \ce{^{*}CHO},
                                \ce{^{*}COH},
                                \ce{^{*}CH2},
                                \ce{^{*}CH2^{*}O},
                                \ce{^{*}CHOH},
                                \ce{^{*}CH3},
                                \ce{^{*}OCH3},
                                \ce{^{*}CH2OH},
                                \ce{^{*}CH4},
                                \ce{^{*}OHCH3} \\ & & \\
        \hline
        C\textsubscript{2}  &   41  &
                                \ce{^{*}C^{*}C},
                                \ce{^{*}CCO},
                                \ce{^{*}CCH},
                                \ce{^{*}CHCO},
                                \ce{^{*}CCHO},
                                \ce{^{*}COCHO},
                                \ce{^{*}CCHOH},
                                \ce{^{*}CCH2},
                                \ce{^{*}CH^{*}CH},
                                \ce{CH2^{*}CO},
                                \ce{^{*}CHCHO},
                                \ce{^{*}CH^{*}COH},
                                \ce{^{*}COCH2O},
                                \ce{^{*}CHO^{*}CHO},
                                \ce{^{*}COHCHO},
                                \ce{^{*}COHCOH},
                                \ce{^{*}CCH3},
                                \ce{^{*}CHCH2},
                                \ce{^{*}COCH3},
                                \ce{^{*}OCHCH2},
                                \ce{^{*}COHCH2},
                                \ce{^{*}CHCHOH},
                                \ce{^{*}CCH2OH},
                                \ce{^{*}CHOCHOH},
                                \ce{^{*}COCH2OH},
                                \ce{^{*}COHCHOH},
                                \ce{^{*}CH2^{*}CH2},
                                \ce{^{*}OCHCH3},
                                \ce{^{*}COHCH3},
                                \ce{^{*}CHOHCH2},
                                \ce{^{*}CHCH2OH},
                                \ce{^{*}OCH2CHOH},
                                \ce{^{*}CHOCH2OH},
                                \ce{^{*}COHCH2OH},
                                \ce{^{*}CHOHCHOH},
                                \ce{^{*}CH2CH3},
                                \ce{^{*}OCH2CH3},
                                \ce{^{*}CHOHCH3},
                                \ce{^{*}CH2CH2OH},
                                \ce{^{*}CHOHCH2OH},
                                \ce{^{*}OHCH2CH3} \\
        \hline
        Nitrogen-based &  24   &
                               \ce{^{*}NH2N(CH3)2},
                                \ce{^{*}ONN(CH3)2},
                                \ce{^{*}OHNNCH3},
                                \ce{^{*}NNCH3},
                                \ce{^{*}ONH},
                                \ce{^{*}NHNH},
                                \ce{^{*}NHN2},
                                \ce{^{*}N^{*}NH},
                                \ce{^{*}ONNO2},
                                \ce{^{*}NO2NO2},
                                \ce{^{*}N^{*}NO},
                                \ce{^{*}N2},
                                \ce{^{*}ONNH2},
                                \ce{^{*}NH2},
                                \ce{^{*}NH3},
                                \ce{^{*}NONH},
                                \ce{^{*}NH},
                                \ce{^{*}NO2},
                                \ce{^{*}NO},
                                \ce{^{*}N},
                                \ce{^{*}NO3},
                                \ce{^{*}OHNH2},
                                \ce{^{*}ONOH},
                                \ce{^{*}CN} \\
        \hline
    \end{tabular}
    \caption{Adsorbates considered in \dataset .}
    \label{tab:adsorbates}
\end{table*}

\section{Train/Test/Validation Splits}

The following adsorbates were reserved for validation subsplits:  *CH, *CHO, *COCH\textsubscript{2}OH, *COH, *NH\textsubscript{2}, *NH\textsubscript{2}N(CH\textsubscript{3})\textsubscript{2}, and *ONOH. Asterisks represent the binding atoms. The following adsorbates were reserved for the test subsilpts:  *CH\textsubscript{2}*CH\textsubscript{2}, *CO, *COHCH\textsubscript{2}, *NHN\textsubscript{2}, *NNCH\textsubscript{3}, *OCHCH\textsubscript{2}, and *ONNO\textsubscript{2}.

\begin{figure*}
    \centering
    \includegraphics[width=0.85\textwidth]{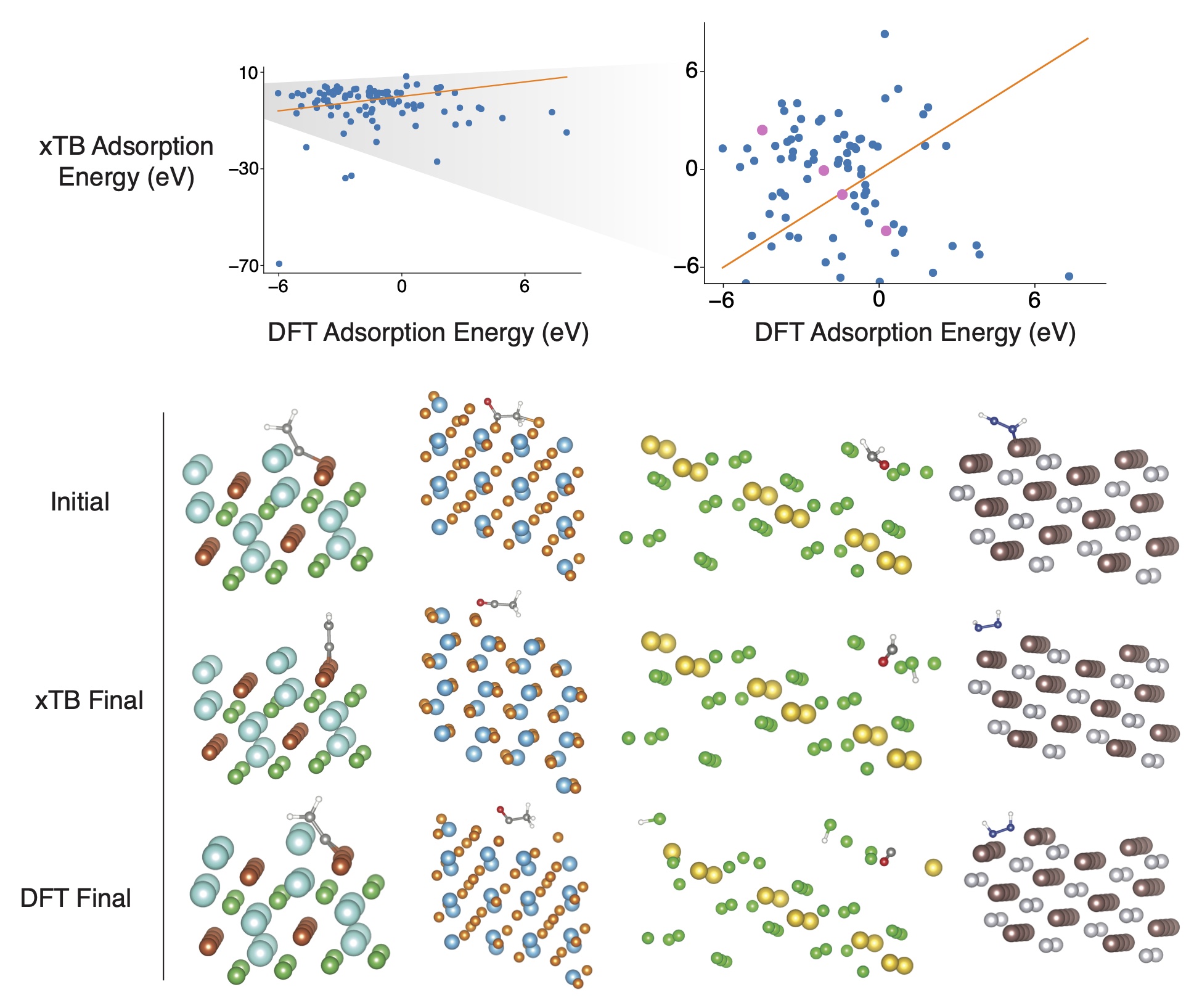}
    \caption{Top: A parity plot comparing xTB adsorption energies with DFT adsorption energies and an inset that limits xTB values to a range similar to that of DFT. Bottom: Initial and final structures corresponding to the pink markers in the plot above organized from left to right.}
    \label{fig:xtb}
\end{figure*}

\section{Tight Binding Baseline}
Obtaining reasonable energies, forces, and relaxed structures from tight binding codes is an enticing possibly because of the low computational cost compared to DFT; however, tight binding calculations on systems for catalysis remain a challenge, as demonstrated by SI Figure \ref{fig:xtb}. We preformed tight binding calculations on 100 random systems from the validation set with extended tight binding (xTB) and the atomic simulation environment (ASE)~\cite{ase_Hjorth_Larsen_2017} interface using the GFN0 parameters~\cite{pracht2019robust}. All xTB calculations were carried out in accordance to our DFT procedures with a few notable differences. For the combined systems, i.e. an adsorbate on a surface, all surface atoms were fixed during the relaxation. Relaxations with xTB featured a BFGS optimizer instead of conjugated gradient, but the convergence criteria remained the same as other DFT calculations, $f_{max}$ of 0.03 eV/\AA{} or a maximum of 200 steps except for adsorbate references where $f_{max}$ was 0.05 eV/\AA{}. Additionally, the surface energies used for the computation of adsorption energies were approximated with single point energies. We did not allow surfaces to relax because of unphysical behavior during optimization, which we likely attribute to periodic boundary conditions (PBCs). We are aware that the xTB code was designed for non-periodic systems and that incorporation of PBCs is an ongoing effort. Overall, the speed of the xTB was impressive and we look forward to future developments related to systems with PBCs.

\section{Additional Data: Rattled \& Molecular Dynamics}

Off-equilibrium data was additionally generated to diversify the structures in the dataset. Two approaches were use to generate this additional data: structural perturbations ("rattled") and molecular dynamics.

\textbf{Rattled}. Structures along the relaxation path way were sampled, perturbed via random atomic position displacements, and evaluated with DFT. For each relaxation, 20\% of the intermediate structures were sampled for rattling. Atomic displacements were sampled from a normal distributions with $\mu=0$ and $\sigma=0.05$. Approximately 30 million single-point calculations were carried out. Upon filtering, $17M$ \gls{S2EF} data points were used for training.

\noindent \textbf{Molecular Dynamics}. Short time-scale molecular dynamics simulations were performed on previously relaxed structures. Simulations took place at 900K for 80 or 320 fs with an integration step size of 2 fs in the NVE ensemble. Approximately 64 million single-point calculations were carried out. Upon filtering, $38M$ \gls{S2EF} data points were used for training.

\noindent
\textbf{Performance of baseline models}.
    We report \gls{S2EF} and \gls{IS2RS} results for SchNet~\cite{schutt2017schnet}
    and DimeNet$++$~\cite{klicpera2020fast} models optimized for force-prediction
    in Table~\ref{tab:rattled_MD_results}.
    Consistent with results in the main paper, we find that DimeNet$++$ outperforms
    SchNet (lower Force MAE, higher Force cosine, higher AFbT).
    Compared to training only on \gls{S2EF} data, training on MD data
    seems to provide a complementary learning signal and leads to better sample
    efficiency -- both DimeNet$++$ and SchNet trained on \gls{S2EF}-$20M$ $+$ MD
    ($58M$ training samples) outperform corresponding models trained on
    \gls{S2EF}-All ($134M$ training samples) as per AFbT.
    Finally, \gls{IS2RS} AFbT seems to correlate better with \gls{S2EF} Force
    cosine than \gls{S2EF} Force MAE, especially when comparing models trained
    on Rattled or MD data.

\setlength{\tabcolsep}{3pt}
\begin{table*}[h!]
    \begin{center}
        \resizebox{\textwidth}{!}{
            \begin{tabular}{ll r ccccc }
                \toprule
                & & & \multicolumn{2}{c}{S2EF Test} & \multicolumn{3}{c}{IS2RS Test} \\
                \cmidrule(l{4pt}r{4pt}){4-5}
                \cmidrule(l{4pt}r{4pt}){6-8}
                Model & Training Data & $\#$ Samples & Force MAE & Force cosine & ADwT & FbT & AFbT \\
                \midrule
                SchNet~\cite{schutt2017schnet} & S2EF-$20M$ & $20M$
                    & $0.0535$ & $0.3006$ & $27.68\%$ & $0.00\%$ & $1.68\%$ \\
                SchNet~\cite{schutt2017schnet} & S2EF-All & $134M$
                    & $0.0490$ & $0.3417$ & $31.78\%$ & $0.00\%$ & $3.38\%$ \\
                SchNet~\cite{schutt2017schnet} & S2EF-$20M$ $+$ Rattled & $37M$
                    & $0.0691$ & $0.3619$ & $36.70\%$ & $0.10\%$ & $5.14\%$ \\
                SchNet~\cite{schutt2017schnet} & S2EF-$20M$ $+$ MD & $58M$
                    & $0.0775$ & $0.3885$ & $41.10\%$ & $0.15\%$ & $8.97\%$ \\
                \midrule
                DimeNet$++$~\cite{klicpera2020directional,klicpera2020fast} & S2EF-$20M$ & $20M$
                    & $0.0509$ & $0.3382$ & $34.37\%$ & $0.00\%$ & $2.67\%$ \\
                DimeNet$++$~\cite{klicpera2020directional,klicpera2020fast} & S2EF-All & $134M$
                    & $0.0357$ & $0.4787$ & $48.91\%$ & $0.25\%$ & $15.17\%$ \\
                DimeNet$++$~\cite{klicpera2020directional,klicpera2020fast} & S2EF-$20M$ $+$ Rattled & $37M$
                    & $0.0658$ & $0.4395$ & $43.94\%$ & $0.05\%$ & $12.51\%$ \\
                DimeNet$++$~\cite{klicpera2020directional,klicpera2020fast} & S2EF-$20M$ $+$ MD & $58M$
                    & $0.0635$ & $0.4644$ & $47.69\%$ & $0.15\%$ & $17.09\%$ \\
                \midrule
                DimeNet$++$~\cite{klicpera2020directional,klicpera2020fast}-large & S2EF-All & $134M$
                    & $0.0313$ & $0.5443$ & $51.67\%$ & $0.40\%$ & $21.74\%$ \\
                \bottomrule
            \end{tabular}}
    \end{center}
    \caption{\gls{S2EF} and IS2RS results of force-only SchNet and DimeNet$++$
        models on \gls{S2EF}, MD, and Rattled data.}
    \label{tab:rattled_MD_results}
\end{table*}
\setlength{\tabcolsep}{1.4pt}

\section{Results on Validation splits}
Full results on the validation splits are shown in Tables \ref{tab:final_S2EF_subsplit_results_val}, \ref{tab:final_structure_results_val}, and \ref{tab:final_energy_subsplit_results_val} for the \gls{S2EF}, \gls{IS2RS}, and \gls{IS2RE} tasks respectively.

\setlength{\tabcolsep}{3pt}
\begin{table*}[h]
    \begin{center}
        \resizebox{\textwidth}{!}{
            \begin{tabular}{l l cccc | cccc  }
            \toprule
            & & \multicolumn{4}{c|}{Energy MAE [eV] $\downarrow$} & \multicolumn{4}{c}{EwT $\uparrow$}  \\
            \cmidrule(l{4pt}r{4pt}){3-6}
            \cmidrule(l{4pt}r{4pt}){7-10}
            & & \multicolumn{8}{c}{Validation}  \\
        Model & Approach & ID &  OOD Ads & OOD Cat & OOD Both & ID &  OOD Ads & OOD Cat & OOD Both \\
            \midrule
            Median baseline & -
                & $1.7466$ & $1.7647$ & $1.7283$ & $1.5640$
                & $0.78\%$ & $0.80\%$ & $0.83\%$ & $0.91\%$ \\
            \midrule
            CGCNN~\cite{xie2018crystal} & Direct
                & $0.6203$ & $0.7426$ & $0.6001$ & $0.6708$
                & $3.36\%$ & $2.11\%$ & $3.53\%$ & $2.29\%$  \\
            SchNet~\cite{schutt2017schnet} & Direct
                & $0.6465$ & $0.7074$ & $0.6475$ & $0.6626$
                & $2.96\%$ & $2.22\%$ & $3.03\%$ & $2.38\%$  \\
            DimeNet$++$~\cite{klicpera2020directional,klicpera2020fast} & Direct
                & $0.5636$ & $0.7127$ & $0.5612$ & $0.6492$
                & $4.25\%$ & $2.48\%$ & $4.40\%$ & $2.56\%$   \\
            \midrule

            SchNet~\cite{schutt2017schnet} & Relaxation
                & $0.7150$ & $0.7395$ & $0.8010$ & $0.8197$
                & $4.03\%$ & $3.09\%$ & $3.87\%$ & $2.72\%$ \\
            SchNet~\cite{schutt2017schnet} -- force-only + energy-only & Relaxation
                & $0.7110$ & $0.7574$ & $0.8316$ & $0.8075$
                & $4.33\%$ & $2.88\%$ & $3.63\%$ & $2.57\%$ \\
            \bottomrule
            \end{tabular}}
    \end{center}
    \caption{Predicting relaxed state energy from initial structure (\gls{IS2RE}) as evaluated by Mean Absolute Error (MAE) of the energies and the percentage of Energies within a Threshold (EwT) of the ground truth energy. Results reported for trained on the All training dataset.}
    \label{tab:final_energy_subsplit_results_val}
\end{table*}
\setlength{\tabcolsep}{1.4pt}

\setlength{\tabcolsep}{3pt}
\begin{table*}[h]
    \begin{center}
        \resizebox{0.8\textwidth}{!}{
            \begin{tabular}{l cccc }
            & \multicolumn{4}{c}{\gls{S2EF} Validation}  \\
            \midrule
            Model  & ID &  OOD Ads & OOD Cat & OOD Both \\
            \midrule
            & \multicolumn{4}{c}{Energy MAE [eV] $\downarrow$} \\
            Median baseline  & $2.0715$ & $2.2275$ & $2.0558$ & $2.3313$ \\
            CGCNN~\cite{xie2018crystal} & $0.5041$ & $0.5986$ & $0.5252$ & $0.7308$ \\
            SchNet~\cite{schutt2017schnet} & $0.4468$ & $0.4973$ & $0.5453$ & $0.7047$ \\
            SchNet~\cite{schutt2017schnet} -- force-only & $34.0183$ & $33.4238$ & $34.2519$ & $38.1693$ \\
            SchNet~\cite{schutt2017schnet} -- energy-only & $0.4011$ & $0.4727$ & $0.5607$ & $0.7165$ \\
            DimeNet$++$~\cite{klicpera2020directional,klicpera2020fast} & $0.4545$ & $0.5093$ & $0.5184$ & $0.6753$  \\
            DimeNet$++$~\cite{klicpera2020directional,klicpera2020fast} -- force-only & $28.2095$ & $28.4266$ & $28.8740$ & $35.0468$  \\
            DimeNet$++$~\cite{klicpera2020directional,klicpera2020fast} -- energy-only & $0.3599$ & $0.4500$ & $0.5412$ & $0.7108$  \\
            DimeNet$++$~\cite{klicpera2020directional,klicpera2020fast}-Large -- force-only & $29.3524$ & $29.4825$ & $29.9799$ & $36.6944$ \\
            \midrule
            & \multicolumn{4}{c}{Force MAE [eV/\AA{}] $\downarrow$} \\
            Median baseline  & $0.0810$ & $0.0799$ & $0.0798$ & $0.0942$ \\
            CGCNN~\cite{xie2018crystal} & $0.0684$ & $0.0746$ & $0.0679$ & $0.0852$ \\
            SchNet~\cite{schutt2017schnet} & $0.0493$ & $0.0574$ & $0.0520$ & $0.0685$ \\
            SchNet~\cite{schutt2017schnet} -- force-only & $0.0442$ & $0.0514$ & $0.0465$ & $0.0618$ \\
            SchNet~\cite{schutt2017schnet} -- energy-only & $0.5810$ & $0.6254$ & $0.5875$ & $0.6562$ \\
            DimeNet$++$~\cite{klicpera2020directional,klicpera2020fast} & $0.0443$ & $0.0508$ & $0.0445$ & $0.0589$  \\
            DimeNet$++$~\cite{klicpera2020directional,klicpera2020fast} -- force-only & $0.0331$ & $0.0366$ & $0.0343$ & $0.0436$  \\
            DimeNet$++$~\cite{klicpera2020directional,klicpera2020fast} -- energy-only & $0.3410$ & $0.3322$ & $0.3425$ & $0.3502$  \\
            DimeNet$++$~\cite{klicpera2020directional,klicpera2020fast}-Large -- force-only & $0.0281$ & $0.0318$ & $0.0315$ & $0.0396$ \\
            \midrule
            & \multicolumn{4}{c}{Force Cosine $\uparrow$} \\
            Median baseline  & $0.0000$ & $0.000$ & $0.000$ & $0.000$ \\
            CGCNN~\cite{xie2018crystal} & $0.1550$ & $0.1320$ & $0.1456$ & $0.1338$ \\
            SchNet~\cite{schutt2017schnet} & $0.3185$ & $0.2862$ & $0.2973$ & $0.2854$ \\
            SchNet~\cite{schutt2017schnet} -- force-only & $0.3604$ & $0.3296$ & $0.3294$ & $0.3266$ \\
            SchNet~\cite{schutt2017schnet} -- energy-only & $0.0841$ & $0.0695$ & $0.0807$ & $0.0699$ \\
            DimeNet$++$~\cite{klicpera2020directional,klicpera2020fast} & $0.3632$ & $0.3401$ & $0.3512$ & $0.3556$  \\
            DimeNet$++$~\cite{klicpera2020directional,klicpera2020fast} -- force-only & $0.4877$ & $0.4747$ & $0.4599$ & $0.4849$  \\
            DimeNet$++$~\cite{klicpera2020directional,klicpera2020fast} -- energy-only & $0.1064$ & $0.0855$ & $0.1043$ & $0.0880$  \\
            DimeNet$++$~\cite{klicpera2020directional,klicpera2020fast}-Large -- force-only & $0.5640$ & $0.5500$ & $0.5106$ & $0.5390$ \\
            \midrule
            & \multicolumn{4}{c}{EFwT $\uparrow$} \\
            Median baseline  & $0.00\%$ & $0.01\%$ & $0.01\%$ & $0.01\%$ \\
            CGCNN~\cite{xie2018crystal} & $0.01\%$ & $0.00\%$ & $0.00\%$ & $0.01\%$ \\
            SchNet~\cite{schutt2017schnet} & $0.13\%$ & $0.00\%$ & $0.10\%$ & $0.00\%$ \\
            SchNet~\cite{schutt2017schnet} -- force-only & $0.00\%$ & $0.00\%$ & $0.00\%$ & $0.00\%$ \\
            SchNet~\cite{schutt2017schnet} -- energy-only & $0.00\%$ & $0.00\%$ & $0.00\%$ & $0.00\%$ \\
            DimeNet$++$~\cite{klicpera2020directional,klicpera2020fast} & $0.09\%$ & $0.00\%$ & $0.09\%$ & $0.00\%$  \\
            DimeNet$++$~\cite{klicpera2020directional,klicpera2020fast} -- force-only & $0.00\%$ & $0.00\%$ & $0.00\%$ & $0.00\%$  \\
            DimeNet$++$~\cite{klicpera2020directional,klicpera2020fast} -- energy-only & $0.00\%$ & $0.00\%$ & $0.00\%$ & $0.00\%$ \\
            DimeNet$++$~\cite{klicpera2020directional,klicpera2020fast}-Large -- force-only & $0.00\%$ & $0.00\%$ & $0.00\%$ & $0.00\%$ \\
         \bottomrule
            \end{tabular}}
    \end{center}
    \caption{Predicting energy and forces from a structure (\gls{S2EF}) as
    evaluated by Mean Absolute Error (MAE) of the energies, force MAE, force cosine,
    and the percentage of Energies and Forces within Threshold (EFwT). Results
    reported for models trained on the entire training dataset (S2EF-All).}
    \label{tab:final_S2EF_subsplit_results_val}
\end{table*}
\setlength{\tabcolsep}{1.4pt}

\setlength{\tabcolsep}{3pt}
\begin{table*}[h]
    \begin{center}
        \resizebox{0.8\textwidth}{!}{
            \begin{tabular}{l cccc }
            & \multicolumn{4}{c}{\gls{IS2RS} Validation}  \\
            \midrule
            Model & ID &  OOD Ads & OOD Cat & OOD Both \\
            \midrule
            & \multicolumn{4}{c}{ADwT $\uparrow$} \\
            IS baseline & $21.18\%$ & $23.49\%$ & $20.25\%$ & $28.29\%$\\
            SchNet~\cite{schutt2017schnet} & $15.53\%$ & $16.57\%$ & $14.50\%$ & $17.29\%$ \\
            SchNet~\cite{schutt2017schnet} -- force-only & $32.41\%$ & $33.33\%$ & $30.02\%$ & $37.48\%$ \\
            DimeNet$++$~\cite{klicpera2020directional,klicpera2020fast} & $30.40\%$ & $30.77\%$ & $29.94\%$ & $34.89\%$ \\
            DimeNet$++$~\cite{klicpera2020directional,klicpera2020fast} -- force-only & $49.05\%$ & $46.91\%$ & $46.54\%$ & $55.23\%$ \\
            \bottomrule
            \end{tabular}}
    \end{center}
    \caption{Predicting relaxed structure from initial structure (\gls{IS2RS})
        as evaluated by Average Distance within Threshold (ADwT).
        All values in percentages, higher is better. Results reported for
        structure to energy-force (S2EF) models trained on the All training
        dataset. The initial structure was used as a naive baseline (IS baseline).
        Note that metrics requiring expensive DFT calculations --
        FbT and AFbT -- are only computed for test splits, not val.}
    \label{tab:final_structure_results_val}
\end{table*}
\setlength{\tabcolsep}{1.4pt}

\section{Changelog}

This section tracks the changes to this document since the original release.

\noindent \textbf{v1}. Intial version.

\noindent \textbf{v2}.
\begin{itemize}[noitemsep,topsep=0pt]
    \item DimeNet\cite{klicpera2020directional} results replaced with DimeNet$++$\cite{klicpera2020fast, klicpera2020directional}.
        DimeNet$++$ is more memory-efficient and performs slightly better.
    \item Force cosine similarity added as an additional \gls{S2EF} metric.
        It correlates better with downstream \gls{IS2RS} AFbT.
    \item 81 systems removed from the original 1.28M systems due to convergence issues later discovered.
        Models were not retrained due to the negligible amount of data ($\sim$0.00675\%).
    \item Rattled/MD data experiments added.
\end{itemize}

\noindent \textbf{v3}. Included additional VASP\cite{Kresse1994, Kresse1996, Kresse1996a, vasp-license, kresse1999ultrasoft} citations.

\noindent \textbf{v4}.
\begin{itemize}[noitemsep,topsep=0pt]
    \item Some systems removed or modified due to subtle convergence and trajectory stitching issues later discovered, affecting a very small proportion of the data. More details can be found at \url{https://github.com/Open-Catalyst-Project/ocp/blob/master/DATASET.md#version-2-feb-2021}. S2EF models were not retrained due to the negligible amount of data affected.
    \item IS2RE models retrained and metrics re-evaluated.
    \item IS2RS - ADwT metrics re-evaluated.
\end{itemize}

\noindent \textbf{v5}.
\begin{itemize}[noitemsep,topsep=0pt]
    \item A bug was resolved in the IS2RE via relaxation approach. Metrics have been updated, results now outperforming the direct-based approaches.
    \item Some systems removed from the validation and test splits due to errors found in their initial placements and/or improper split classification ($\sim$0.167\% IS2RE, $\sim$0.043\% S2EF). 
    \item S2EF, IS2RE, and IS2RS metrics were re-evaluated with the updated spits.
\end{itemize}
\end{document}